\documentclass[12pt,journal,draftcls,a4paper,onecolumn]{IEEEtran}
\usepackage[nomarkers,nofiglist,notablist]{endfloat} % places all figures at the end of document
\usepackage{dsfont}
\usepackage{amsmath}
\usepackage{amssymb}
\usepackage{amsfonts}
\usepackage{graphicx}
\usepackage{amsmath}
\usepackage[all,poly]{xy}
\usepackage{multirow}
\usepackage{algorithm}
\usepackage{algorithmic}
\usepackage{color}
\usepackage[noadjust]{cite}
\usepackage{subfigure}
\usepackage{tikz,pgf}
\usepackage{url}
\usetikzlibrary{arrows,automata}
\usepackage{verbatim}
\usetikzlibrary{positioning,shapes.geometric}

\newcommand{\figwidth}{\columnwidth}
\newcommand{\bone}{\boldsymbol{1}}
\newcommand{\bzero}{\boldsymbol{0}}

\newcommand{\bmu}{\boldsymbol{\mu}}

\newcommand{\beps}{\boldsymbol{\epsilon}}
\newcommand{\bthe}{\boldsymbol{\theta}}
\newcommand{\bThe}{\boldsymbol{\Theta}}
\newcommand{\bsig}{\boldsymbol{\sigma}}
\newcommand{\bSig}{\boldsymbol{\Sigma}}
\newcommand{\bgam}{\boldsymbol{\gamma}}

\newcommand{\bLam}{\boldsymbol{\Lambda}}

\newcommand{\bstau}{\boldsymbol{\tau}}
\newcommand{\bsswh}{\textbf{SWH}}
\newcommand{\bsPu}{\boldsymbol{P_u}}

\def\bse{{\boldsymbol{e}}}

\def\bss{{\boldsymbol{s}}}

\def\bsx{{\boldsymbol{x}}}
\def\bsy{{\boldsymbol{y}}}

\def\bsD{{\boldsymbol{D}}}

\def\bsF{{\boldsymbol{F}}}

\def\bsY{{\boldsymbol{Y}}}

\def\calG{{\mathcal{G}}}

\def\calI{{\mathcal{I}}}

\def\calN{{\mathcal{N}}}

\newcounter{algo}
\renewcommand{\thealgo}{\arabic{algo}}

\title{Bayesian Estimation of Smooth Altimetric Parameters: Application to Conventional and Delay/Doppler Altimetry}
\author{Abderrahim Halimi$^{\,1}$\thanks{(1) A. Halimi and H. Snoussi are with the Institut Charles Delaunay (CNRS), Universit\' e de technologie de Troyes, France (e-mail: \{Abderrahim.Halimi, Hichem.Snoussi\}@utt.fr).}, Corinne Mailhes$^{\,2}$, Jean-Yves Tourneret$^{\,2}$
\thanks{(2) C.  Mailhes and J.-Y. Tourneret are  with the University of Toulouse, IRIT/INP-ENSEEIHT/T\'eSA, Toulouse, France (e-mail: \{Corinne.Mailhes,
Jean-Yves.Tourneret\}@enseeiht.fr).} and Hichem Snoussi$^{\,1}$ }

\begin{document}

\maketitle

\begin{abstract}
This paper proposes a new Bayesian strategy for the smooth estimation of altimetric parameters. The altimetric signal is assumed to be corrupted by a thermal and speckle noise distributed according to an independent and non identically Gaussian distribution. We introduce a prior enforcing a smooth temporal evolution of the altimetric parameters which improves their physical interpretation. The posterior distribution of the resulting model is optimized using a gradient descent algorithm which allows us to compute the maximum a posteriori estimator of the unknown model parameters. This algorithm presents a low computational cost which is suitable for real time applications. The proposed Bayesian strategy and the corresponding estimation algorithm are validated on both synthetic and real data associated with conventional and delay/Doppler altimetry. The analysis of real Jason-$2$ and Cryosat-$2$ waveforms shows an improvement in parameter estimation when compared to the state-of-the-art estimation algorithms.
\end{abstract}


%%%%%%%%%%%%%%%%%%%%%%%%%%%%%%%%%%%%%%%%%%%%%%%%%%%
%%%%%%%%%%%%%%%%%%%%%%%%%%%%%%%%%%%%%%%%%%%%%%%%%%%
%%%%%%%%%%%%%%%%%%%%%%%%%%%%%%%%%%%%%%%%%%%%%%%%%%%
\section{Introduction} \label{sec:Introduction}
A satellite altimeter is a nadir viewing radar that emits regular pulses and records the travel time, the magnitude and the shape of each return signal after reflection on the Earth's surface. This reflected echo provides information about many parameters such as the range between the satellite and the observed scene (denoted by $\tau$), the significant wave height (denoted by SWH) and the wind speed (related to the echo's amplitude $P_u$). However, altimetric waveforms are corrupted by  speckle noise that reduces the quality of the retrieved geophysical parameters. Therefore, many recent studies and missions have been focusing on  improving the quality of these estimated parameters by reducing the noise effect. We distinguish between two approaches that have been focusing on  improving the altimetric technology or improving the processing of the available data. The first approach is based on reducing the measurement noise of conventional altimeters (such as Topex and Poseidon-3) by increasing the number of observations (looks). This can be achieved by keeping the same conventional technology and improving the altimeter characteristics (as for AltiKa \cite{Vincent2006,Richard2007,Raney2011}), or by using a new delay/Doppler processing \cite{Raney1998} that results in a different echo's shape \cite{Wingham2004,HalimiTGRS2014,HalimiTGRS2014b}. The second approach processes the available data  by using improved models for the altimetric echoes \cite{Brown1977,Amarouche2004,HalimiTGRS2012,Halimi2013} or improved estimation algorithms \cite{Maus1998,Sandwell2005,Severini2008}.

Following this second approach, the goal of this paper is to elaborate a new Bayesian model and its estimation algorithm accounting for the smooth evolution of  altimetric signals.
Usual methods for parameter estimation are based on  maximum-likelihood \cite{Dumont1985} or  weighted \cite{Maus1998,Halimi2013Eusipco} and unweighted least squares (LS) approaches \cite{Amarouche2004,HalimiTGRS2014,HalimiTGRS2014b}.
Most existing algorithms consider altimetric echoes independently and estimate one set of parameters per echo. However, it is well known that adjacent echoes share similar characteristics because of their physical nature. This correlation has been considered in  \cite{Maus1998} which processed a set of echoes by imposing a spline evolution for the epoch parameter $\tau$. However, this method was developed by fixing the significant wave height for the whole set of echoes.  In \cite{Sandwell2005}, the authors proposed a three-step algorithm that reduces the range noise  by first estimating three altimetric parameters (SWH, $\tau$ and $P_u$), then filtering the estimated SWH using a Gaussian filter and finally re-estimating the epoch $\tau$ while fixing SWH. This last algorithm is flexible since it does not enforce the parameter SWH to be constant. However, it requires to use a sequential estimator whose computational cost can be prohibitive. Moreover, the algorithm of \cite{Sandwell2005} requires the characterization of a Gaussian filter for SWH which has been chosen empirically. Also, it is interesting to note that the algorithms developed in \cite{Maus1998,Sandwell2005}  have shown interesting properties especially in reducing the variance of the estimated altimetric parameters which encourages the consideration of the echo's correlation to improve parameter estimation.

The first contribution of this paper is the elaboration of a hierarchical Bayesian model that allows smooth estimation of altimetric parameters by considering the correlation between successive altimetric echoes. Each altimetric echo is assumed to be corrupted by an additive, independent and non identically Gaussian noise \cite{Puig2009} whose mean is the thermal noise. This noise model generalizes the independent and identically distributed (i.i.d.) Gaussian noise that is generally assumed when considering an un-weighted LS approach \cite{Amarouche2004,HalimiTGRS2014}. The proposed approach allows the estimation of the noise parameters from the data which brings additional robustness compared to existing algorithms and allows the processing of different kinds of data.
The parameters of interest of the proposed model (i.e., SWH, $\tau$, $P_u$)
are assigned priors enforcing a smooth evolution between consecutive signals which improves their estimation. This prior is defined from the discrete Laplacian of the different parameters. It has shown increasing interest for many problems such as image deconvolution \cite{Campisi2007,Molina2006}, hyperspectral unmixing \cite{Sigurdsson2014} and medical imaging \cite{Chaari2013}.
Note that the proposed Bayesian hierarchy is generic in the sense that it can be applied to both conventional altimetry (CA) and delay/Doppler altimetry (DDA).

The second contribution of this paper is the derivation of an estimation algorithm for the altimetric and noise parameters associated with the proposed hierarchical Bayesian model. The minimum mean square error (MMSE) and maximum a posteriori (MAP) estimators of these parameters cannot be easily computed from the obtained joint posterior. This problem has been solved in \cite{Halimi2015ICASSP} by drawing samples distributed according to this posterior using Markov chain Monte Carlo (MCMC) methods. More precisely, an Hamiltonian Monte Carlo (HMC) algorithm was used for its good mixing properties for high-dimensional vectors \cite{Brooks2011}. The HMC algorithm provided good estimation results and will be briefly recalled in this paper. However, the resulting HMC-based estimators have a high computational complexity which can be penalizing for real time applications. In this paper, we propose an algorithm that significantly reduces this computational cost by using a gradient descent approach. More precisely, a coordinate descent algorithm
\cite{Bertsekas1995,Sigurdsson2014} is used to sequentially update the noise and altimetric parameters. This algorithm uses a natural gradient approach \cite{Amari1998}, weighting the gradient direction by the Fisher information matrix and leading to a fast convergence\footnote{This approach is also known as Fisher scoring \cite{Green1984,HalimiTGRS2012}. It updates the parameters in a Riemannian space resulting in a fast convergence to a local minimum of the cost function of interest \cite{Girolami2011,Pereyra2013}.}.
The proposed Bayesian model and estimation algorithm are validated using synthetic and real echoes acquired during the Jason-$2$ and Cryosat-$2$ missions.  The obtained results are very promising and confirm the accuracy of the proposed Bayesian model and its associated algorithm.

The paper is organized as follows. Section
\ref{sec:Altimetric_models} presents the different models used for
CA and DDA. The proposed hierarchical Bayesian model and its estimation algorithm are introduced in Sections \ref{sec:Hierarchical_Bayesian_model} and \ref{sec:Estimation_algorithms}. Section \ref{sec:Simulations} is devoted to testing and  validating the proposed technique using synthetic data with controlled ground truth. Section \ref{sec:Results_on_real_data} shows results obtained using real data resulting from Jason-2 and Cryosat-2 missions.
Conclusions and future work are finally reported in Section
\ref{sec:Conclusions}.

%%%%%%%%%%%%%%%%%%%%%%%%%%%%%%%%%%%%%%%%%%%%%%%%%%%
%%%%%%%%%%%%%%%%%%%%%%%%%%%%%%%%%%%%%%%%%%%%%%%%%%%
%%%%%%%%%%%%%%%%%%%%%%%%%%%%%%%%%%%%%%%%%%%%%%%%%%%
%\newpage
\section{Altimetric models} \label{sec:Altimetric_models}
This section introduces the altimetric models used for CA and DDA that are characterized by three altimetric parameters: the significant wave height $\textrm{SWH}$, the epoch $\tau$ and the amplitude $P_u$.

\subsection{Conventional and Brown models} \label{subsec:Conventional_model}
In CA, the mean power of the altimetric echo $s(t)$ is expressed by a convolution of three terms that are the probability density function (PDF) of the
heights of the specular scatterers, the point
target response (PTR) of the radar\footnote{Instead of the considered squared cardinal sine PTR, a measured PTR could also be used in CA as in  \cite{AmaroucheOSTST2014,BoyOSTST2014}.} and the flat surface
impulse response (FSIR) as follows \cite{Brown1977,Hayne1980,Amarouche2004}
\begin{equation} s(t) =
\textrm{FSIR}(t)*\textrm{PDF}(t) *\textrm{PTR}_T(t)
 \label{eqt:Double_convolution_CA}
\end{equation}
\vspace{-0.15cm} with \vspace{-0.15cm}
\begin{equation}
\textrm{FSIR}(t) = P_u \exp{\left[-\frac{4 c}{\gamma h}
\left(t-\tau_s \right) \right]} U\left(t-\tau_s \right)
\label{eqt:FSSR_bessel_CA_xi0}
\end{equation}
\begin{equation}
\textrm{PDF}(t) = \frac{1}{\sqrt{2 \pi} \sigma_s}
\exp{\left( - \frac{t^2}{2 \sigma_s^2 } \right) } \label{eqt:PDF}
\end{equation}
\begin{equation}
 \textrm{PTR}_T(t) =   {\left| \frac{
\sin{\left( \pi \frac{t}{T} \right) } }{\pi \frac{t}{T}}
\right|}^2 \label{eqt:PTR_T}
\end{equation}
where $\sigma_s = \frac{\textrm{SWH}} {2 c}$, $t$ is the time, $\tau_s$ is the epoch expressed in
seconds, $c$ is the speed of light, $T$ is the time resolution, $\gamma$ is an antenna beamwidth parameter, $h$ is the minimum satellite-surface distance and $U(.)$ denotes the Heaviside function. This model has been considered for parameter estimation in many studies such as \cite{HalimiTGRS2014,Halimi2013Eusipco}. However, the numerical computation of the double convolution \eqref{eqt:Double_convolution_CA} leads to a high computational cost when processing a large amount of altimetric data. Therefore, a simplified analytical model known as the Brown model (BM) is often considered.
%\subsection{Brown model} \label{subsec:Brown_model}
%As explained previously, t
The BM is obtained by assuming a centered Gaussian approximation (with a standard-deviation equals to $\sigma_p$) for the squared cardinal sine $\textrm{PTR}_T$, leading to \cite{Brown1977,Amarouche2004}
\begin{equation}
s(t)= \frac{P_u}{2}\left[ 1+ \textrm{erf} \left( \frac{t-\tau_s -\alpha
\sigma _{c}^{2}}{\sqrt{2}\sigma _{c}}\right) \right]
\exp\left[-\alpha \left( t-\tau_s -\frac{\alpha \sigma
_{c}^{2}}{2}\right)
\right]   \label{eqt:Brown_model}
\end{equation}
where
\begin{equation}
\qquad \sigma _{c}^{2}=\left(\frac{\textrm{SWH}}{2c}
\right)^{2}+\sigma _{p}^{2}\label{modelecont2}
\end{equation}
and where  $\textrm{erf}\left( t\right) =\frac{2}{\sqrt{\pi }}
\int_{0}^{t}e^{-z^{2}}\,dz$ stands for the Gaussian error
function, $\alpha$ and
$\sigma_{p}^{2}$ being two known parameters (depending on the
satellite and on the Gaussian approximation of the PTR). The model \eqref{eqt:Brown_model} is clearly more attractive than \eqref{eqt:Double_convolution_CA} for designing simple estimation algorithms.

\subsection{Delay/Doppler altimetry} \label{subsec:Delay_Doppler_model}
DDA aims at reducing the measurement noise and increasing the along-track resolution in comparison with CA. The mean power of a delay/Doppler echo can be expressed as
the convolution of three terms: the time/frequency FSIR, the PDF of
the heights of the specular scatterers  and the
time/frequency PTR as follows \cite{Puig2009,Phalippou2011,HalimiTGRS2014,HalimiTGRS2014b}
\begin{equation}
P(t,f) =
\textrm{FSIR}(t,f)  *\textrm{PDF}(t) *\textrm{PTR}(t,f)
 \label{eqt:Double_convolution}
\end{equation}
\vspace{-0.2cm} with \vspace{-0.2cm}
\begin{equation}
\textrm{PTR}(t,f) = \textrm{PTR}_T(t) \textrm{PTR}_F(f),
\;\; \textrm{PTR}_F(f) =   {\left| \frac{ \sin{\left( \pi
\frac{f}{F} \right) } }{\pi \frac{f}{F}} \right|}^2
 \label{eqt:PTRF}
\end{equation}
where $F$ is the frequency resolution.
The FSIR is given by  \cite{HalimiTGRS2014}
\begin{gather}
\textrm{FSIR}(t,q) =   \frac{P_u}{\pi}
\exp{\left[-\frac{4 c}{\gamma h}
\left(t-\tau_s\right) \right]} \, U\left(t-\tau_s\right) \nonumber\\
\times \left[ \phi_{t,q+1}^{}(\tau_s)-\phi_{t,q}^{}(\tau_s)
\right]
 \label{eqt:Doppler_pol_xi0_2}
\end{gather}
\vspace{-0.15cm} with \vspace{-0.25cm}
\begin{equation}
\phi_{t,q}(\tau_s) =
\operatorname{Re}\left[\textrm{arctan}\left(
\frac{y_q}{\sqrt{\rho^2(t-\tau_s)-y_q^2}} \right)\right]
\end{equation}
\vspace{-0.05cm}where $q=1, \cdots,Q$, $Q$ being the
number of Doppler beams ($Q=64$ for Cryosat-$2$), $\rho(t)=\sqrt{h c t}$ is the
radius of the propagation circles (iso-range circles), $y_q = \frac{h \lambda}{2
v_s} f_q$ is the ordinate of the $q^{\textrm{th}}$ Doppler
beam, $f_q = \left(q-32.5\right) F$ is the $q^{\textrm{th}}$
Doppler frequency ($32.5$ is introduced to obtain a central beam in the case where $Q=64$),
$v_s$ is the satellite velocity, $\lambda$ is the wavelength
and $\operatorname{Re}(x)$ denotes the real part of the
complex number $x$ (the interested reader is invited to consult \cite{HalimiTGRS2014} for more details about this formula). The double convolution \eqref{eqt:Double_convolution} leads to a 2-D delay/Doppler map (DDM). The analysis of the DDA  waveform is based on a multi-look echo obtained by applying Doppler processing (slant range correction and
multi-looking) to the DDM (see \cite{Raney1998,Wingham2004}). The resulting multi-look delay/Doppler signal can be written
\begin{equation}
s(t) = \sum_{q=1}^{Q} P\left( t-\delta t_q,f_q \right)
\label{eqt:Multi_look_echo}
\end{equation}
where $\delta t_q$ is the delay compensation expressed in
seconds.
Note finally that the discrete altimetric echo (for CA, BM, and DDA) is gathered in the vector $\bss = \left(s_1, \cdots, s_K\right)^T$, where $K=128$ gates and $s_k = s\left(k T \right)$.

\subsection{Noise model} \label{subsec:Noise_model}
The altimetric echoes are corrupted by a speckle noise whose influence is reduced by averaging, on-board the satellite, a sequence of $L$ consecutive echoes. This operation
reduces the noise variance by $\sqrt{L}$ when assuming pulse-to-pulse statistical independence between the averaged echoes\footnote{This assumption is true for CA echoes while it is not valid for DDA echoes as explained in \cite{Wingham2004,Halimi2013Eusipco}.}. Considering the central limit theorem and using the fact that the averaging is conducted on a large number of echoes, the resulting noise sequence can be approximated by a Gaussian distribution. This approximation has been successfully adopted in the altimetric community \cite{Puig2009,Germain2006,Halimi2013Eusipco} and is implicitly used in the well known least squares estimation algorithms  \cite{Puig2010,HalimiTGRS2014,Maus1998,Sandwell2005}. Therefore, this paper considers that the altimetric waveform is corrupted by an additive Gaussian noise as follows
\begin{equation}
\bsy_m = \bss_m\left( \bThe_{m} \right) + \bse_m, \textrm{  with   } \bse_m \sim \calN \left( \mu_m \bone_K, \bSig_m \right)
\label{eqt:Observation_model}
\end{equation}
where $\bsy_m$ (resp. $\bss_m$) is a ($K \times 1$) vector representing the $m$th observed echo (resp. altimetric model), $\bThe_{m}=\left[\theta_1(m),\theta_2(m),\theta_3(m)\right] = \left[\textrm{SWH}(m),\tau(m),P_u(m)\right] $ is a ($1 \times 3$) vector containing the $3$ altimetric parameters $\textrm{SWH}, \tau, P_u$ for the $m$th echo, $\bse_m$ is an independent and non identically Gaussian noise with mean vector $\mu_m \bone_K$ representing the thermal noise (where $\bone_K$ is a $K \times 1$ vector of $1$), and diagonal covariance matrix $\bSig_m = \textrm{diag} \left(\bsig^2_m \right)$ with $\bsig^2_m =\left(\sigma^2_{m1},\cdots,\sigma^2_{mK}\right)^T$ a ($K\times 1$) vector containing the noise variances of the $m$th echo\footnote{Note that the noise variance varies with respect to the time instant $k$ and the observed echo $m$.}. Note that the thermal noise parameter is sometimes represented by an additive constant that is added to the altimetric model $\bss_m\left( \bThe_{m} \right)$. In the proposed model, this parameter has been included in the noise sequence $\bse_m$ of \eqref{eqt:Observation_model} where it represents the noise mean $\mu_m$.
The next section introduces the Bayesian model associated with a set of $M$ successive echoes considered in this paper.

%%%%%%%%%%%%%%%%%%%%%%%%%%%%%%%%%%%%%%%%%%%%%%%%%%%
%%%%%%%%%%%%%%%%%%%%%%%%%%%%%%%%%%%%%%%%%%%%%%%%%%%
%%%%%%%%%%%%%%%%%%%%%%%%%%%%%%%%%%%%%%%%%%%%%%%%%%%
%\clearpage
\section{Hierarchical Bayesian model} \label{sec:Hierarchical_Bayesian_model}
This section introduces a hierarchical Bayesian model for estimating the parameters of $M$ successive altimetric echoes. The Bayesian approach assigns prior distributions to the unknown parameters summarizing the prior knowledge about these parameters.   More precisely, if $f\left(\bsx \right)$  denotes the prior distribution assigned to the parameter $\bsx$, the Bayesian approach computes the posterior distribution of $\bsx$ using Bayes rule
\begin{equation}
f(\bsx|\bsY) = \frac{f(\bsY|\bsx) f(\bsx)}{f(\bsY)} \propto f(\bsY|\bsx) f(\bsx) \label{eqt:Bayes}
\end{equation}
where $\propto$ means ``proportional to'' and $f(\bsY|\bsx)$ is the likelihood of the observation vector $\bsY$.
The vector $\bsx$ is then estimated from this distribution by computing its mean (MMSE estimator) or its maximum (MAP estimator).
The following sections introduce the likelihood and the prior distributions considered in this paper. The unknown parameters of the proposed model include the thermal noise represented by an  ($M \times 1$) vector $\bmu$, the ($K \times M$) matrix $\bLam=\left[\bsig^2_1,\cdots,\bsig^2_M \right]$ containing the noise variances associated with the considered $M$ echoes, and the ($M \times 3$) matrix gathering the $3$ altimetric parameters (SWH, $\tau$ and $P_u$) of the $M$ echoes.

\subsection{Likelihood} \label{subsec:Likelihood}
The observation model defined in \eqref{eqt:Observation_model} and the Gaussian properties of the noise sequence $\bse_m$ yield
\begin{equation}
f(\bsy_{m}|\bThe_{m}, \mu_m, \bSig_m) \propto {\left( \frac{1}{ \prod_{k=1}^{K} \sigma^2_{mk} }\right)
}^{\frac{1}{2}} \exp
\left\lbrace -\frac{1}{2}  \bsx_m^T  \bSig_m^{-1} \bsx_m  \right\rbrace  \label{eqt:likelihood}
\end{equation}
where  $\bsx_m = \bsy_{m}-\bss_m -  \mu_m \bone_K $, and $\bsx_m\left( \bThe_{m} \right)$ (resp. $\bss_m\left( \bThe_{m} \right)$) has been denoted by $\bsx_m$ (resp. $\bss_m$) for brevity. Assuming independence between the observations leads to
\begin{equation}
f(\bsY|\bThe, \bmu, \bLam) \propto \prod_{m=1}^{M}{f(\bsy_{m}|\bThe_{m}, \mu_m, \bSig_m)  }
\label{eqt:likelihood_globale}
\end{equation}
where $\bThe =\left[\bthe_1,\bthe_2,\bthe_3\right]$ is an ($M\times 3$) matrix containing the altimetric parameters of $M$ echoes, $\bmu = \left(\mu_1,\cdots,\mu_M\right)^T$ is an ($M \times 1$) vector containing the noise means and $\bLam=\left[\bsig^2_1,\cdots,\bsig^2_M \right]$ is a ($K \times M$) matrix containing the noise variances.

\subsection{Priors for the altimetric parameters} \label{subsec:Prior_for_altimetric_parameters}
The altimetric parameters are assigned priors enforcing a smooth evolution for successive waveforms. The considered priors constrain the derivatives of the altimetric parameters to have small values. This can be done by assigning a Gaussian distribution to the second derivative of the altimetric parameter $\bthe_{i}$ as follows
\begin{equation}
f(\bthe_{i}|\epsilon^2_i) \propto
\left(\frac{1}{\epsilon_i^2}\right)^{M/2} \exp{\left(-\frac{1}{2 \epsilon_i^2}  \left\| \bsD \bthe_{i}\right\|^2 \right)}    \label{eqt:prior_parameters}
\end{equation}
for $i\in \left\lbrace 1,\cdots,3 \right\rbrace$, where  $\epsilon^2_i$ is an hyperparameter, $ ||\cdot|| $ denotes the standard $l_2 $ norm such that $ ||\bsx||^2 = \bsx^T \bsx$, and $\bsD$ is the discrete Laplacian operator. This prior has been referred to as simultaneous autoregression (SAR) or conditional autoregression (CAR) in the image deconvolution context \cite{Campisi2007,Molina2006}. It has also been used for different problems  such as spectral unmixing of hyperspectral images \cite{Sigurdsson2014}, medical imaging applications \cite{Chaari2013} or spectroscopy applications \cite{Mazet2011}. The main originality of this work is to consider this prior for the parameters of altimetric waveforms.
%Note, moreover, that we constrained the estimated parameter to belong to the set $\left[\ell_i, u_i \right]$ of realistic values as follows $\textrm{SWH} \in \left[0,50\right]$ m, $\tau \in \left[5, 70\right]$ gates and $P_u>0$.

\subsection{Prior for the noise parameters} \label{subsec:Prior_for_the_Noise_parameters}
The noise parameters introduced in \eqref{eqt:Observation_model} are also estimated by the proposed estimation algorithm. The absence of knowledge about the noise means (which represent the thermal noise) and covariances can be issued by choosing a non informative priors for these parameters. More precisely, we consider the following conjugate Gaussian prior with a large variance for the noise mean
 %Jeffreys prior for the noise mean
\begin{equation}
f(\bmu|\psi^2) \sim \calN \left(\bzero_M, \psi^2 \mathbf{I}_M \right)
\label{eqt:Nt_prior}
\end{equation}
%\propto \bone_M
where $\mathbf{I}_M$ is the ($M \times M$) identity matrix, $\bzero_M$ is an ($M \times 1$) vector of $0$ and  $\psi^2$ is a fixed large variance ($\psi^2=10^2$ in our simulations).
%where $\bone_M$ is an ($M \times 1$) vector of $1$.
Considering the noise covariance matrix, we have first to note that in the case of speckle noise, these variances depend on the echo's shape at each instant, i.e., the variances change from one echo to the other. However, we are generally interested in   parameters estimated with a frequency of $1$ Hz, i.e., we consider that the physical parameters do not change significantly in $1$ second. This assumption means that the shapes of successive echoes acquired within a time interval of $1$ s do not change significantly and thus that the noise variances are almost the same for these echoes. Therefore, we consider that the noise variance is constant for each $r=20$ successive echoes\footnote{The pulse repetition frequency of Jason-$2$ is $2$ kHz. The echoes are generally averaged to deliver a $20$ Hz data, i.e., it delivers $20$ echoes for each second of data.}. This means that $ \sigma^2_{(n-1)r+1,k}= \cdots = \sigma^2_{nr,k}$ for $n \in \left\lbrace1,\cdots,N \right\rbrace$, with $N=\frac{M}{r}$  (note that the general case is obtained by considering $r=1$).
After assuming prior independence between the noise variances $\sigma^2_{nr,k}$, the non informative scale prior of $\bLam$ is defined as
\begin{equation}
f\left(\bLam\right) = \prod_{n=1}^{N} {\prod_{k=1}^{K} { \frac{1}{\sigma^2_{nr,k}}   \mathbb{I}_{\mathbb{R}^+} \left(\sigma^2_{nr,k}\right) }}  \label{eqt:priorSigma2}
\end{equation}
where $\mathbb{I}_{\mathbb{R}^+} (.)$ is the indicator of the set $\mathbb{R}^+$.

\subsection{Hyperparameter priors} \label{subsec:Prior_for_the_hyperparameters}
The hyperparameters $\epsilon_i^2, i \in \left\lbrace1,\cdots,3 \right\rbrace$ are assigned independent inverse gamma distributions as follows
\begin{equation}
\epsilon_i^2\ \sim   \calI \calG \left(a_i, b_i\right)   \label{eqt:prior_epsilon}
\end{equation}
where $\sim$ means ``is distributed according to'' and $a_i$, $b_i$ are fixed hyperparameters that depend on the dynamic magnitude of the $i$th altimetric parameter \cite{Mazet2011}.

\subsection{Marginalized posterior distribution} \label{subsec:Posterior_distribution}
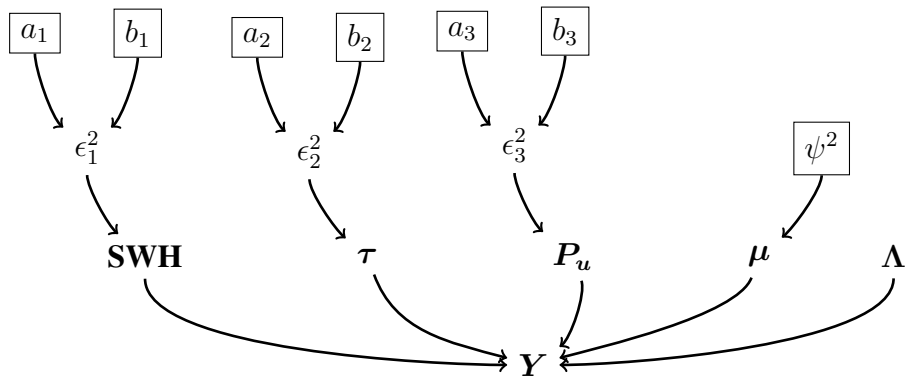
\begin{figure}[h!]
\centering
\begin{tikzpicture}
 nodes %
\node[text centered] (Y) {$\bsY$};
\node[above =1.  of Y, text centered] (LL) {};
\node[left  =1.75  of LL, text centered] (tau) {$\bstau$};
\node[left  =2  of tau, text centered] (swh) {$\bsswh$};
\node[right =2  of tau, text centered] (pu) {$\bsPu$};
%\node[right =2  of pu, text centered] (e) {$\bse$};
\node[above =1.  of swh, text centered] (Tswh) {$$};
\node[above =1.  of tau, text centered] (Ttau) {$$};
\node[above =1.  of pu, text centered] (TPu) {$$};
%\node[above =1.25  of e, text centered] (Te) {$$};
\node[right =0.2  of Tswh, text centered] (eps1) {$ $};
\node[left  =0.3  of Tswh, text centered] (m1) {$\epsilon_1^2$};
\node[right =0.3  of Ttau, text centered] (eps2) {$ $};
\node[left  =0.3  of Ttau, text centered] (m2) {$\epsilon_2^2$};
\node[right =0.8  of TPu, text centered] (eps3) {$ $};
\node[left  =0.3  of TPu, text centered] (m3) {$\epsilon_3^2$};
\node[right =3.5  of pu, text centered] (sig) {$\bLam$};
\node[right =1.75  of pu, text centered] (Nt) {$\bmu$};
\node[above =1.  of Nt, text centered] (Tmu) {$$};
\node[draw, rectangle,right  =0.3  of Tmu, text centered] (psi) {$\psi^2$};
\node[above =1.  of m1, text centered] (Teps1) {$$};
\node[draw, rectangle, left  =0.2  of Teps1, text centered] (a1) {$a_1$};
\node[draw, rectangle, right =0.2  of Teps1, text centered] (b1) {$b_1$};
\node[above =1.  of m2, text centered] (Teps2) {$$};
\node[draw, rectangle, left  =0.2  of Teps2, text centered] (a2) {$a_2$};
\node[draw, rectangle, right =0.2  of Teps2, text centered] (b2) {$b_2$};
\node[above =1.  of m3, text centered] (Teps3) {$$};
\node[draw, rectangle, left  =0.2  of Teps3, text centered] (a3) {$a_3$};
\node[draw, rectangle, right =0.2  of Teps3, text centered] (b3) {$b_3$};

 edges %
\draw[->, line width= 1] (tau) to  [out=290,in=165, looseness=1] (Y);
\draw[->, line width= 1] (swh) to  [out=270,in=180, looseness=0.5] (Y);
\draw[->, line width= 1] (pu) to  [out=290,in=35, looseness=0.5] (Y);
\draw[->, line width= 1] (m1) to  [out=270,in=140, looseness=0.5] (swh);
\draw[->, line width= 1] (m2) to  [out=270,in=140, looseness=0.5] (tau);
\draw[->, line width= 1] (m3) to  [out=270,in=140, looseness=0.5] (pu);
\draw[->, line width= 1] (psi) to [out=270,in=40, looseness=0.5]  (Nt);
\draw[->, line width= 1] (a1) to  [out=270,in=140, looseness=0.5] (m1);
\draw[->, line width= 1] (a2) to  [out=270,in=140, looseness=0.5] (m2);
\draw[->, line width= 1] (a3) to  [out=270,in=140, looseness=0.5] (m3);
\draw[->, line width= 1] (b1) to  [out=270,in=40, looseness=0.5] (m1);
\draw[->, line width= 1] (b2) to  [out=270,in=40, looseness=0.5] (m2);
\draw[->, line width= 1] (b3) to  [out=270,in=40, looseness=0.5] (m3);
\draw[->, line width= 1] (sig) to  [out=270,in=0, looseness=0.5] (Y);
\draw[->, line width= 1] (Nt) to  [out=250,in=15, looseness=0.5] (Y);
\end{tikzpicture}
\caption{DAG for the parameter and hyperparameter priors (the fixed parameters appear in boxes). }
\label{fig:DAG}
\end{figure}

The proposed Bayesian model includes the parameters $\bThe, \bmu, \bLam$ and the hyperparameters $\epsilon_i^2, i \in \left\lbrace1,\cdots,3 \right\rbrace$ which is summarized in the directed acyclic graph (DAG) displayed in Fig. \ref{fig:DAG}. The joint posterior distribution of this Bayesian model can be computed from the following hierarchical structure
\begin{equation}
f\left(\bThe, \bmu, \bLam, \beps | \bsY \right)  \propto f(\bsY|\bThe, \bmu, \bLam)  f\left(\bThe, \bmu, \bLam, \beps  \right)
\label{eqt:Joint_Posterior}
\end{equation}
with
\begin{eqnarray}
f\left(\bThe, \bmu, \bLam, \beps \right) & = &   f\left(\bmu|\psi^2\right)  f\left(\bLam  \right)  \prod_{i=1}^{3}{f(\bthe_{i}|\epsilon^2_i) f(\epsilon^2_i|a_i,b_i)}
\label{eqt:Prior}
\end{eqnarray}
where we have assumed a priori independence between the altimetric parameters and hyperparameters. This distribution can be marginalized with respect to (w.r.t.) $\beps$ as follows
\begin{eqnarray}
f\left(\bThe, \bmu, \bLam  | \bsY \right) & \propto & f(\bsY|\bThe, \bmu, \bLam) f\left(\bmu|\psi^2\right)  f\left(\bLam  \right)  \prod_{i=1}^{3}{\left[\int {f(\bthe_{i}|\epsilon^2_i) f(\epsilon^2_i|a_i,b_i)} d\epsilon^2_i\right]}  \nonumber \\
 &\propto &f(\bsY|\bThe, \bmu, \bLam) f\left(\bmu|\psi^2\right)  f\left(\bLam  \right) \prod_{i=1}^{3}{ f\left(\bthe_{i}|a_i,b_i \right)  }
\label{eqt:Marg_Posterior1}
\end{eqnarray}
where
\begin{equation}
f\left(\bthe_{i}|a_i,b_i   \right)  \propto    \left( \frac{ \left\| \bsD \bthe_{i}\right\|^2}{2}+ b_i    \right)^{-a_i - M/2}.
\label{eqt:Prior_theta_margin}
\end{equation}
The advantage of this marginalization is to get rid of the hyperparameter vector $\beps$ and thus to make  the estimation procedure more robust. However, even with this simplification, the MMSE and MAP estimators associated with the marginalized posterior \eqref{eqt:Marg_Posterior1} are not easy to determine mainly because of the nonlinearities relating the noiseless altimetric waveform $\bss_m$ and the parameters of interest ($\tau$ and SWH). In \cite{Halimi2015ICASSP}, an MCMC algorithm was proposed to sample the joint posterior distribution of a similar Bayesian model. The generated samples were then used to compute the MMSE parameter estimators (see Appendix \ref{app:Hybrid_Gibbs_algorithm} for more details about this algorithm). However, this MCMC algorithm has a high computational complexity which  can be penalizing for real time applications. In this paper, we propose an alternative based on an optimization technique maximizing the marginalized posterior \eqref{eqt:Marg_Posterior1} w.r.t. the parameters of interest (or equivalently, minimizing the negative log-posterior $ - \textrm{log} [ p(\bThe, \bmu, \bLam | \bsY)]$ defined in \eqref{eqt:Marg_Posterior1}). The mode of the posterior distribution  \eqref{eqt:Marg_Posterior1} is classically referred to as marginalized MAP estimator of ($\bThe, \bmu, \bLam$).
%After removing unnecessary constants, the negative log-posterior reduces to the following cost function
%\begin{eqnarray}
%\mathcal{C} \left(\bThe, \bmu, \bLam  \right)   & =  &
%\left(\frac{r}{2}+1\right) \sum_{k=1}^{K}{    \sum_{n=1}^{N}{\log \sigma_{nr,k}^2}    } + \sum_{m=1}^{M}{\frac{\bsx_m^T  \bSig_m^{-1} \bsx_m}{2}   } \nonumber \\
%& + & \sum_{i=1}^{3}{ \left( a_i +\frac{M}{2}\right) \log\left( \frac{ \left\| \bsD \bthe_{i}\right\|^2}{2}+ b_i    \right) } + \sum_{m=1}^{M}{\frac{\mu_m^2}{2\psi^2}}.
%\label{eqt:Cost_function}
%\end{eqnarray}
%%with $\bthe_{i} \in \left[\ell_i, u_i \right]^M, \forall i \in \left\lbrace1,\cdots,3\right\rbrace$.

%%%%%%%%%%%%%%%%%%%%%%%%%%%%%%%%%%%%%%%%%%%%%%%%%%%
%%%%%%%%%%%%%%%%%%%%%%%%%%%%%%%%%%%%%%%%%%%%%%%%%%%
%%%%%%%%%%%%%%%%%%%%%%%%%%%%%%%%%%%%%%%%%%%%%%%%%%%
%\clearpage
\section{Coordinate descent algorithm} \label{sec:Estimation_algorithms}

%%%%%%%%%%%%%%%%%%%%%%%%%%%%%%%%%%%%%%%%%%%%%%%%%%%
%\newpage
%\subsection{Gradient descent algorithm} \label{subsec:Gradient_descent_algorithm}
This section describes a gradient descent algorithm maximizing the marginalized posterior \eqref{eqt:Marg_Posterior1} w.r.t. the noise and altimetric parameters, i.e., determining the marginal MAP estimator of ($\bThe, \bmu, \bLam$). As explained before, maximizing \eqref{eqt:Marg_Posterior1} w.r.t. ($\bThe, \bmu, \bLam$)  reduces to minimizing the negative log-posterior defined as (after removing unnecessary constants)
\begin{eqnarray}
\mathcal{C} \left(\bThe, \bmu, \bLam  \right)   & =  &
\left(\frac{r}{2}+1\right) \sum_{k=1}^{K}{    \sum_{n=1}^{N}{\log \sigma_{nr,k}^2}    } + \sum_{m=1}^{M}{\frac{\bsx_m^T  \bSig_m^{-1} \bsx_m}{2}   } \nonumber \\
& + & \sum_{i=1}^{3}{ \left( a_i +\frac{M}{2}\right) \log\left( \frac{ \left\| \bsD \bthe_{i}\right\|^2}{2}+ b_i    \right) } + \sum_{m=1}^{M}{\frac{\mu_m^2}{2\psi^2}}.
\label{eqt:Cost_function}
\end{eqnarray}
Because of the large number of parameters in ($\bThe, \bmu, \bLam$), we propose a coordinate descent algorithm \cite{Bertsekas1995,Sigurdsson2014} that sequentially updates the different parameters. More precisely, the algorithm estimates the altimetric and noise parameters iteratively.  In each step, the parameters of interest are estimated while holding the other parameters fixed. This process is repeated until the algorithm has converged to a local minimum of the cost function $\mathcal{C}(\bThe, \bmu, \bLam)$. The main steps of the resulting algorithm are described in Algo. \ref{alg:Gradient_descent_algorithm}. The next sections describe the sub-optimization procedures maximizing the cost function $\mathcal{C}(\bThe, \bmu, \bLam)$ w.r.t. the altimetric  and  noise parameters. A convergence diagnosis is also provided to study the convergence of the proposed approach.
\begin{algorithm}
\caption{Gradient descent algorithm} \label{alg:Gradient_descent_algorithm}
\begin{algorithmic}[1]
       \STATE \underline{Initialization}
       \STATE Initialize parameters $\bthe_{i}^{(0)},$ for $i \in  \left\lbrace 1,\cdots,3 \right\rbrace$, $\bmu^{(0)}$, $\bLam^{(0)}$ and $t=1$
       \STATE conv$=0$,
       \STATE \underline{Parameter update}
       \WHILE{conv$=0$}
               \STATE Update $\bthe_{i}^{(t)},$ for $i \in  \left\lbrace 1,\cdots,3 \right\rbrace$  according to \eqref{eqt:natural_gradient}
               \STATE Update $\bmu^{(t)}$ according to \eqref{eqt:posterior_mu_m_mean}
               \STATE Update $\bLam^{(t)}$ according to \eqref{eqt:sigma2_mean}
               \STATE Set conv$=1$ if the convergence criteria are satisfied
               \STATE t = t + 1
       \ENDWHILE
			%\STATE Compute the MMSE estimate of $\bLam^{(t)}$ using \eqref{eqt:sigma2_mode_chgt}
\end{algorithmic}
\end{algorithm}

\subsubsection{Natural gradient for altimetric parameters} \label{subsubsec:Natural_gradient_for_altimetric_parameters}
In order to ensure a fast estimation of the altimetric parameters, we propose to use the natural gradient algorithm to minimize $\mathcal{C}$ w.r.t. $\bThe$ \cite{Amari1998}. This gradient algorithm is also known as  Fisher scoring approach since it weights the gradient direction by the Fisher information matrix of the parameters \cite{Green1984,HalimiTGRS2012}. The resulting descent algorithm updates the parameters in a Riemannian space
resulting in a fast convergence to a local minimum of the objective function \cite{Girolami2011,Pereyra2013}.  Denote as
$\bgam = \left(\bthe_1^T,\bthe_2^T,\bthe_3^T\right)^T$ the ($3M \times 1$) vector gathering the altimetric parameters. The proposed algorithm updates this vector as follows
\begin{equation}
\bgam^{(t+1)} =  \bgam^{(t)} - \bsF^{-1}\left(\bgam^{(t)} \right)   \nabla\mathcal{C}\left(\bgam^{(t)} \right)
\label{eqt:natural_gradient}
\end{equation}
where $\nabla\mathcal{C}\left(\bgam^{(t)} \right)$ denotes the gradient of the cost function $\mathcal{C}$ given by
\begin{equation}
\nabla\mathcal{C} = \frac{\partial \mathcal{C}}{\partial \bgam}
= \left[\left(\frac{\partial \mathcal{C}}{\partial \bthe_1}\right)^T,\left(\frac{\partial \mathcal{C}}{\partial \bthe_2}\right)^T,\left(\frac{\partial \mathcal{C}}{\partial \bthe_3}\right)^T \right]^T
\label{eqt:gradient_C}
\end{equation}
with
\begin{equation}
\frac{\partial \mathcal{C}}{\partial \bthe_i} = -\left(\frac{\partial \bss_1^T}{\partial \theta_{1i}}   \bSig_1^{-1} \bsx_1,\cdots, \frac{\partial \bss_M^T}{\partial \theta_{Mi}}   \bSig_M^{-1} \bsx_M \right)^T + \left( a_i +\frac{M}{2}\right)  \left( \frac{\bsD^T\bsD  \bthe_i}{\frac{ \left\| \bsD \bthe_{i}\right\|^2}{2}+ b_i} \right).
\label{eqt:gradient_C}
\end{equation}
The Fisher information matrix $\bsF\left(\bgam^{(t)}\right)$ can be decomposed as the following block matrix
\begin{equation}
\bsF = \left[\begin{array}{ccc}
         \bsF_{1} & \bsF_{12} & \bsF_{13} \\
         \bsF_{12} & \bsF_{2} & \bsF_{23} \\
         \bsF_{13} & \bsF_{23} & \bsF_{3}
       \end{array}\right]
\label{eqt:fisher_matrix}
\end{equation}
with
\begin{eqnarray}
\bsF_{ij} &= &\textrm{diag}\left(\frac{\partial \bss_1^T}{\partial \theta_{1i}}   \bSig_1^{-1} \frac{\partial \bss_1}{\partial \theta_{1j}},\cdots, \frac{\partial \bss_M^T}{\partial \theta_{Mi}}   \bSig_M^{-1} \frac{\partial \bss_M}{\partial \theta_{Mj}}       \right)
\label{eqt:fisher_matrix_part_ij} \nonumber \\
\bsF_{i} &= &\bsF_{ii} + \left( a_i +\frac{M}{2}\right)  \left[ \frac{
\left(\frac{ \left\| \bsD \bthe_{i}\right\|^2}{2}+ b_i\right) \bsD^T\bsD - \left(\bsD^T\bsD  \bthe_i\right) \left(\bsD^T\bsD  \bthe_i\right)^T
}{\left(\frac{ \left\| \bsD \bthe_{i}\right\|^2}{2}+ b_i\right)^2} \right].
\label{eqt:fisher_matrix_part_i}
\end{eqnarray}
Note that the derivatives required to compute the elements of $\bsF$ are the same as those required to compute  $\nabla\mathcal{C}$. Therefore, the only additional computation w.r.t. a steepest descent approach is the matrix inversion in \eqref{eqt:natural_gradient}. Note also that this matrix inversion is faster than a new evaluation of the altimetric model necessary for a new update of $\bgam$ in a steepest descent algorithm. Therefore, it is better to add this inversion and reduce the number of iterations, instead of using a simple steepest descent approach that usually requires  a lot of iterations to converge resulting in a high computational cost.
%Note finally that the inequality constraints ($\bthe_i^{(t)} \in \left[q_l,q_u\right]$) are satisfied by replacing the sample that violates the constraints at each iteration by the value of the violated bound, as in \cite{BioucasDias2012}.

\subsubsection{Updating the noise parameters} \label{subsubsec:Updating_the_noise_parameters}
Maximizing the marginalized posterior distribution \eqref{eqt:Marg_Posterior1} w.r.t. the noise parameters (means and variances) can be achieved by considering a coordinate descent algorithm. This approach reduces to maximizing the conditional distributions associated with each noise parameter. Using \eqref{eqt:likelihood_globale} and \eqref{eqt:Nt_prior}, it can be easily shown that the conditional distribution of the noise mean is the following Gaussian distribution
\begin{equation}
\mu_m | \bsy_{m}, \bThe_{m}, \bSig_m,\psi^2  \sim   \calN \left(\overline{\mu_m }
, \frac{1}{\psi^{-2}+ \sum_{k=1}^{K} \sigma_{mk}^{-2} } \right)
\label{eqt:posterior_mu_m}
\end{equation}
with
\begin{equation}
\overline{\mu_m }= \frac{     \sum_{k=1}^{K} \frac{y_{mk} - s_{mk} }{\sigma_{mk}^{2}}  }{\psi^{-2}+ \sum_{k=1}^{K} \sigma_{mk}^{-2}}.
\label{eqt:posterior_mu_m_mean}
\end{equation}
Therefore, the parameter vector $\bmu$ can be updated using \eqref{eqt:posterior_mu_m_mean} which is the maximum of the Gaussian distribution \eqref{eqt:posterior_mu_m}.
Similarly, it can be shown using \eqref{eqt:likelihood_globale} and \eqref{eqt:priorSigma2} that
\begin{equation}
f\left(\bLam | \bsY, \bThe, \bmu \right) = \prod_{n=1}^{N} {\prod_{k=1}^{K} {f\left(\sigma_{nr,k}^2 | \bsY_{:k}, \bThe_{k:}, \bmu\right)}}  \label{eqt:posteriorSigma2}
\end{equation}
and that $\sigma_{nr,k}^2 | \bsY_{:k}, \bThe_{k:}, \bmu$  is distributed according to the following inverse-gamma distribution
\begin{equation}
\sigma_{nr,k}^2 | \bsY_{:k}, \bThe_{k:}, \bmu  \sim   \calI \calG \left(\frac{r}{2}, \beta\right)
\label{eqt:posterior_sigma_nk}
\end{equation}
with $\beta = \sum_{m=(n-1)r+1}^{n r}{\frac{x_{mk}^2}{2}}$.  Thus, the mode of the inverse gamma distribution \eqref{eqt:posterior_sigma_nk} is
\begin{eqnarray}
\overline{\sigma_{nk}^2} = \frac{\beta}{\frac{r}{2}+1}. \label{eqt:sigma2_mean}
\end{eqnarray}
%This mode is used to update the noise variance in Algo. \eqref{alg:Gradient_descent_algorithm}. However, because the inverse gamma distribution is asymmetric, its mode $\overline{\sigma_{nk}^2}$ is not equal to its mean $\widetilde{\sigma_{nk}^2}$. Therefore, and in order to obtain the MMSE estimate for this parameter that has shown better results in simulations, we compute the following transformation at the end of the algorithm //////
%\begin{eqnarray}
%\widetilde{\sigma_{nk}^2} = \overline{\sigma_{nk}^2}  \frac{r+2}{r-2}.
%\label{eqt:sigma2_mode_chgt}
%\end{eqnarray}
%Note finally that the obtained results for both estimators are shown in the simulation sections.

\subsubsection{Stopping criteria} \label{subsubsec:Convergence_diagnosis}
Algo. \ref{alg:Gradient_descent_algorithm} is an iterative algorithm that requires the definition of some stopping criteria. In this paper, we have considered three criteria and the algorithm is stopped if one of them is satisfied.  The first criterion compares the new value of the cost function to the previous one and stops the algorithm if the relative error between these two values is smaller than a given threshold, i.e.,
\begin{equation}
| \mathcal{C}^{(t+1)}-\mathcal{C}^{(t)} | \leq  \xi_1 \mathcal{C}^{(t)}
\label{eqt:criteria1}
\end{equation}
where $|.|$ denotes the absolute value. The second criterion evaluates the new parameter values and stops the algorithm if the following condition is satisfied
\begin{equation}
 \left\| \bgam^{(t+1)}-\bgam^{(t)}   \right\| \leq  \xi_2 \left( \left\|\bgam^{(t)}\right\|  + \xi_2\right).
\label{eqt:criteria2}
\end{equation}
The last criterion is based on a maximum number of iterations $T_{\textrm{max}}$. Note that the proposed algorithm can be used when considering different altimetric models. Indeed, it just needs the definition of the altimetric model and its derivatives w.r.t. the altimetric parameters (the altimetric model can be given by a closed form expression as for BM, or by a numerical computation as for CA and DDA). This model flexibility is highlighted in Section \ref{sec:Results_on_real_data} that considers the BM, CA and DDA models for processing real data.

It is also interesting to note that the coordinate gradient algorithm converges if the minimum of the cost function \eqref{eqt:Cost_function} w.r.t. ($\bThe, \bmu, \bLam$) along each coordinate is uniquely reached \cite{Bertsekas1995}. This is easily checked for the mean vector $\bmu$  since the posterior distribution \eqref{eqt:Marg_Posterior1}  viewed as a function of $\bmu$ is proportional to the Gaussian distribution \eqref{eqt:posterior_mu_m} that has a unique mode. Similarly,  the posterior distribution \eqref{eqt:Marg_Posterior1}  viewed as a function of $\sigma_{nr,k}^2$ is proportional to the inverse gamma distribution \eqref{eqt:posterior_sigma_nk}  that has a unique mode. Considering the altimetric parameters, the convexity  of the cost function cannot be theoretically demonstrated. However, in our simulations  obtained with synthetic and real data (see Sections \ref{sec:Simulations} and \ref{sec:Results_on_real_data}) the proposed algorithm always converges to the actual parameters which tends to confirm this convexity. Note finally that other gradient descent based approaches have been successfully used in the literature to only estimate the altimetric parameters \cite{Dumont1985,Amarouche2004} (without imposing parameter smoothness nor estimating the noise statistics).

%
% most altimetric algorithms are based on a gradient descent algorithm \cite{Dumont1985,Amarouche2004} or geometrical approach \cite{HalimiTGRS2012} that implicitly assumes the convexity of the cost function w.r.t. these parameters. Therefore, the proposed approach converges as long as the actual algorithms do. However, the proposed approach should have better convergence characteristics since the cost function includes additional prior information on the altimetric parameters that regularizes more the criterion.}
%%%%%%%%%%%%%%%%%%%%%%%%%%%%%%%%%%%%%%%%%%%%%%%%%%%
%%%%%%%%%%%%%%%%%%%%%%%%%%%%%%%%%%%%%%%%%%%%%%%%%%%
%%%%%%%%%%%%%%%%%%%%%%%%%%%%%%%%%%%%%%%%%%%%%%%%%%%
%\clearpage
\section{Validation on synthetic data} \label{sec:Simulations}
This section first introduces the criteria used to evaluate the performances of the proposed estimation algorithm. The second part defines the effective number of looks used to exploit the estimated noise covariance. The third part of this section compares the performance of the proposed algorithm with those obtained with state-of-the-art approaches for synthetic data with controlled ground truth.
 %The last part of this section analyzes simulation results obtained with real data acquired during Jason-2 and Cryosat-2 missions.

%Second, it analyzes the performance of the proposed strategy and compares it to the state-of-the-art LS algorithm when considering synthetic BM echoes.

\subsection{Evaluation criteria} \label{subsec:Evaluation_criteria}
For synthetic echoes, the true parameters are supposed to be known and can be used to evaluate the estimation performance of an algorithm. More precisely, the bias and  standard-deviation (STD) of the estimator $\widehat{\theta_i}$ are defined as follows
\begin{equation}
 \textrm{Bias}\left( \widehat{\bthe_i}\right)  =
\frac{1}{M}\sum_{m=1}^{M} \left[{\widehat{\theta_i} {(m)} -
 \theta_i{(m)}}\right]
 \label{eqt:bias}
\end{equation}
 and
\begin{equation}
 \textrm{STD}\left( \widehat{\bthe_i}\right)   =   \sqrt{
\frac{1}{M}\sum_{m=1}^{M} {\left[ \widehat{\theta_i} {(m)}
- \theta_i {(m)} \right]}^2} \label{eqt:STD}
\end{equation}
for $ i \in \left\lbrace1,\cdots,3 \right\rbrace$, where $\theta_i{(m)}$ (resp. $\widehat{\theta_i}{(m)}$ ) is the true (resp. estimated) parameter for the $m$th
waveform and $M$ is the number of simulated waveforms.

\subsection{Effective number of looks} \label{subsec:Effective_number_of_looks}
The proposed algorithm estimates the noise covariance $\hat{\bLam}$ which is used to deduce the effective number of looks (ENL) \cite{Wingham2004,Halimi2013Eusipco}. This number represents the equivalent number of independent echoes that were averaged to reduce the noise  and is defined by
\begin{equation}
N_{\textrm{eff}}  (n,k) = \frac{
E^2\left[y(k)\right]}{E\left\lbrace \left[ y(k)
-E\left(y(k)\right) \right]^2\right\rbrace} = \frac{\left[1/r \sum_{m=(n-1)r+1}^{nr}{y_m(k)}\right]^2    }{\widehat{\sigma^2}_{nr,k}}
\label{eqt:N_eff}
\end{equation}
for $k \in \left\lbrace 1,\cdots,K\right\rbrace$ and $n \in \left\lbrace 1,\cdots,N\right\rbrace$. Note that $N_{\textrm{eff}}(n,k)$ differs from one temporal gate to another and thus depends on $k$.  In order to avoid this dependency, the following averaged number of looks is considered in this paper
\begin{equation}
\textrm{ENL}(n) = \frac{1}{K}\sum_{k=1}^{K}{N_{\textrm{eff}}(n,k)}
\label{eqt:ENL}
\end{equation}
for $n \in \left\lbrace 1,\cdots,N\right\rbrace$. In the case of synthetic data, this estimation will be compared to the number $L$ of averaged echoes in terms of bias and standard-deviation.

\subsection{Simulation results on synthetic data} \label{subsec:Simulation_results_on_synthetic_data}
\begin{figure}[h!]
\centering \subfigure[SWH]{\includegraphics[width=0.75\figwidth,height=6cm]{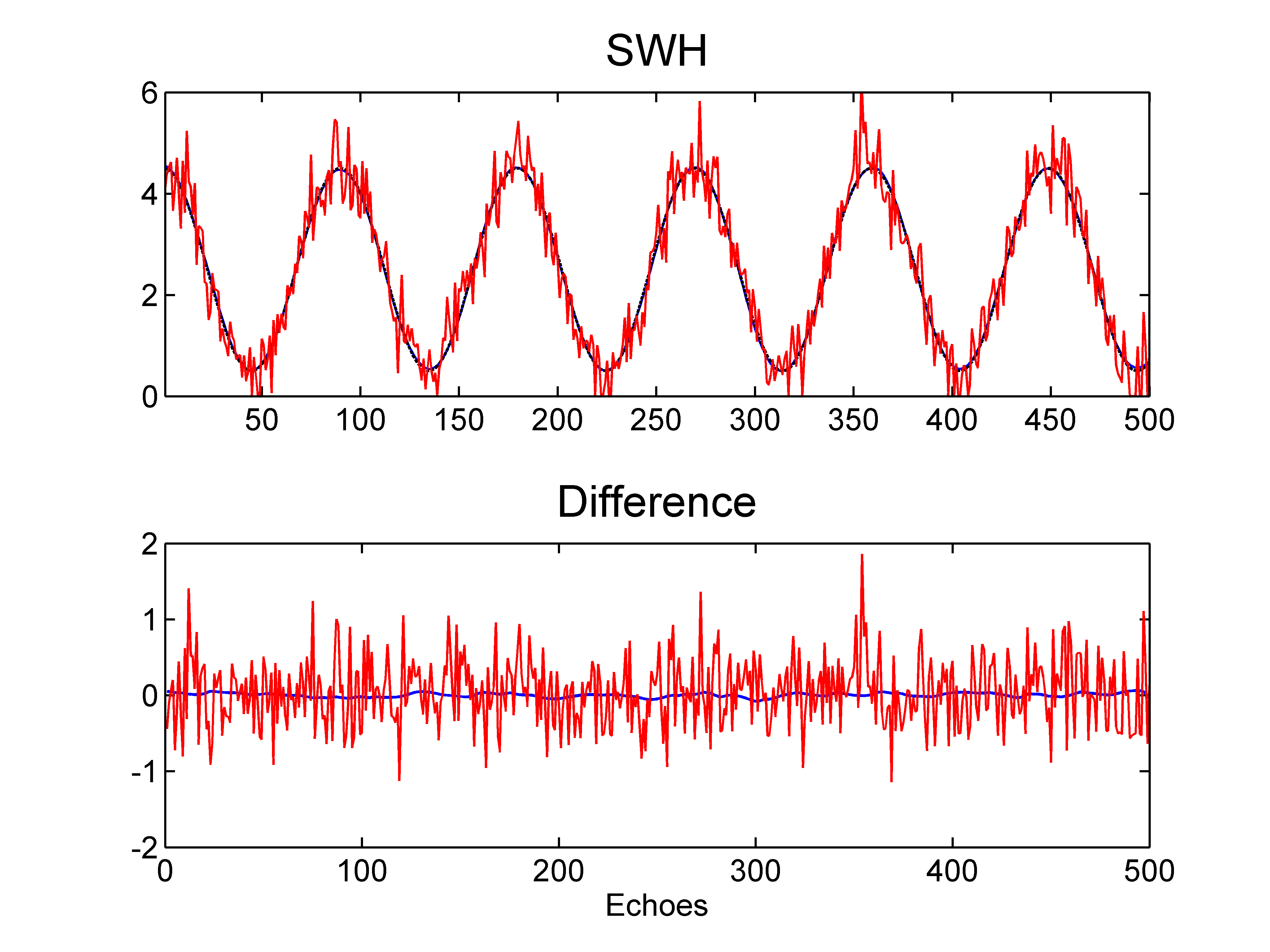}}
\subfigure[$\tau$]{\includegraphics[width=0.75\figwidth,height=6cm]{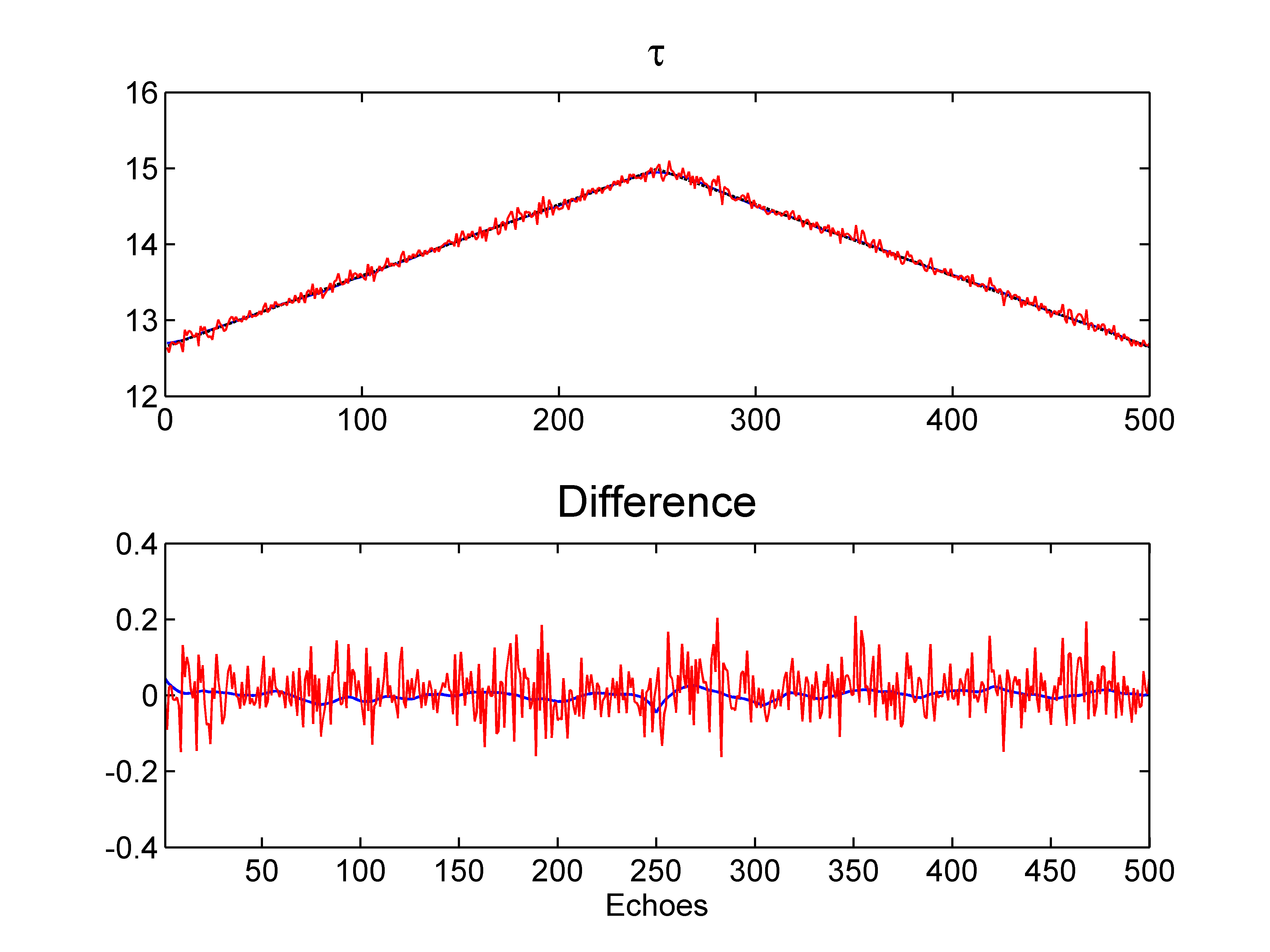}}
\subfigure[$P_u$]{\includegraphics[width=0.75\figwidth,height=6cm]{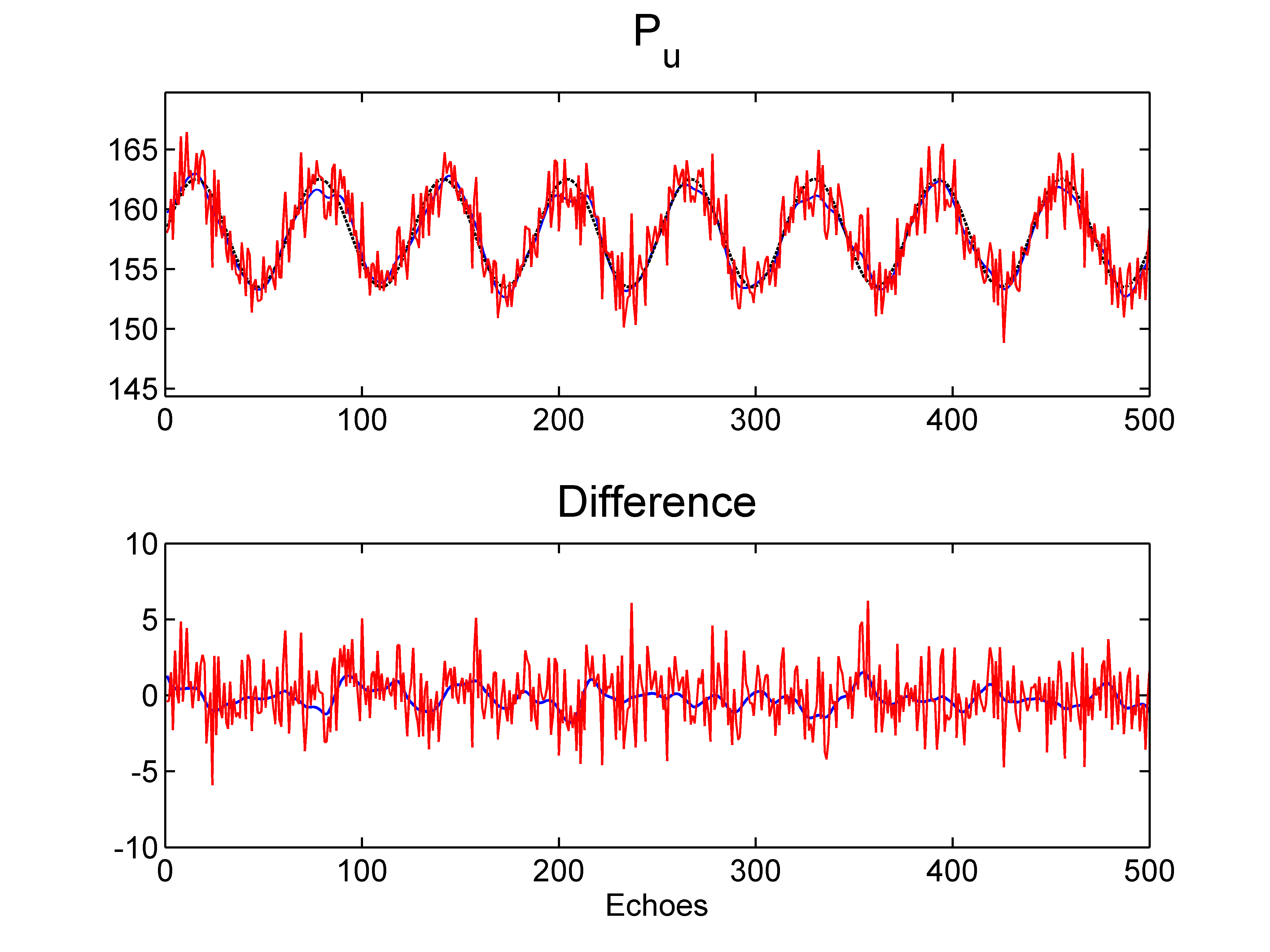}}
\caption{Actual (black line) and estimated parameter using the LS algorithm (red line)  and the proposed CD-BM algorithm (blue line). (a) SWH, (b) $\tau$ and (c) $P_u$. The bottom subfigures show  the difference between the estimated and real parameters.} \label{fig:Synth_estimation_evol_Params_500_CDABr}
\end{figure}

The proposed strategy (denoted by CD for coordinate descent) is first investigated when considering $M=500$ correlated altimetric echoes generated according to the BM defined in \eqref{eqt:Brown_model}. The correlation between successive echoes is introduced by considering a smooth evolution of the altimetric parameters. More precisely, we have considered a dynamic evolution of the altimetric parameters defined as follows   $\textrm{SWH}(m) = 2.5+2 \cos(0.07 m)$, $\tau(m) = 27+0.02m $ if $m<250$ and $\tau(m) = 32-0.02m $ if $m\geq250$, and $P_u(m) = 158+0.05 \sin(0.1 m)$, where $m$ denotes the echo number. The synthetic echoes are then generated by plugging the values of the altimetric parameters in the BM and by corrupting the resulting echo by a thermal noise $\mu_m=0.025$, $\forall m$ and a speckle noise resulting from the averaging of $L=90$ echoes. The proposed CD strategy is compared to the state-of-the-art  LS algorithm described in \cite{Amarouche2004,HalimiTGRS2014,HalimiTGRS2012} and to the Bayesian approach HMC  \cite{Halimi2015ICASSP}, detailed in Appendix \ref{app:Hybrid_Gibbs_algorithm}. Table  \ref{tab:Bias_STD_Synth} reports the obtained biases and STDs when estimating the three parameters of interest and ENL when it is possible. The considered algorithms are denoted by their names (CD, HMC and LS) followed by the considered altimetric model (BM, CA and DDA). For instance, CD-BM represents the CD algorithm when considering BM echoes. Table  \ref{tab:Bias_STD_Synth} shows reduced biases for $\tau$ and SWH when considering the Bayesian CD  and HMC methods. Indeed, these two algorithms estimate the non-identically distributed noise   while the LS estimator only estimates the thermal noise  and assumes an i.i.d. speckle noise. Thus, the Bayesian algorithms involve a more precise model and provide a lower bias for the altimetric parameters (except for $P_u$). Note that all algorithms provide a good estimate of the thermal noise $\bmu$ as shown in Table  \ref{tab:Bias_STD_Synth}.  Moreover, from an STD point of view, both Bayesian CD and HMC approaches lead to lower STDs than when using an LS-based method since they are using the smooth prior defined in \eqref{eqt:prior_parameters}.
Note that CD-BM improves the LS-BM STDs by a factor of $16$ for SWH, $5$ for $\tau$ and $3$ for $P_u$.
Table \ref{tab:Bias_STD_Synth} also highlights the additional interest of both Bayesian algorithms that are able to estimate the number of averaged independent echoes with a good precision (small bias and STD). Last but not least,  Table \ref{tab:Bias_STD_Synth}  gives the averaged computational times of the three algorithms, for a Matlab implementation on a 2.4-GHz Intel-i$7$ PC. As expected, the HMC algorithm is highly computationally expensive, due to its sampling procedure, while the proposed CD algorithm is the fastest one, $3$ times faster than the LS one.
Results similar to  Table \ref{tab:Bias_STD_Synth} have been found when considering CA and DDA models but these results are not provided in this Section for brevity (these models are investigated in the next Section when considering Cryosat-$2$ echoes).
Finally, in the case of synthetic data, the proposed CD approach outperforms the classical LS one, both in terms of bias and STD of the estimated parameters but also in terms of computational cost, allowing at the same time the estimation of additional noise parameters. Figure \ref{fig:Synth_estimation_evol_Params_500_CDABr} illustrates the clear improvement of the proposed algorithm when compared to the LS one, in terms of bias and STD. Sub-figures (a-top), (c-top) and (e-top) present the actual parameter values (black line) and the estimated ones by considering the LS (red line) and the CD (blue line, almost confused with the black one) algorithms for $500$ echoes. When plotting the difference between the actual and estimated parameters in sub-figures (a-bottom), (b-bottom) and (c-bottom), the interest of the proposed CD strategy is clearly highlighted.

%%%\begin{figure}[h!]
%%%\centering
%%%\includegraphics[width=0.7\figwidth]{images/Synth_estimation_evol_SWH_500_CDABr.png}
%%%\caption{Actual (black line) and estimated SWH parameter using the LS algorithm (red line)  and the proposed CDA-BM algorithm (blue line).} \label{fig:Synth_estimation_evol_SWH_500_CDABr}
%%%\end{figure}
%%%\begin{figure}[h!]
%%%\centering
%%%\includegraphics[width=0.7\figwidth]{images/Synth_estimation_evol_TAU_500_CDABr.png}
%%%\caption{Actual (black line) and estimated $\tau$ parameter using the LS algorithm (red line)  and the proposed CDA-BM algorithm (blue line).} \label{fig:Synth_estimation_evol_TAU_500_CDABr}
%%%\end{figure}
%%%\begin{figure}[h!]
%%%\centering
%%%\includegraphics[width=0.75\figwidth]{images/Synth_estimation_evol_PU_500_CDABr.png}
%%%\caption{Actual (black line) and estimated $P_u$ parameter using the LS algorithm (red line)  and the proposed CDA-BM algorithm (blue line).} \label{fig:Synth_estimation_evol_PU_500_CDABr}
%%%\end{figure}
%%\begin{figure}[h!]
%%\centering
%%\includegraphics[width=0.75\figwidth]{images/Synth_estimation_evol_Looks_500_CDABr.png}
%%\caption{Estimated number of looks (L) using the proposed CDA-BM algorithm.} \label{fig:Synth_estimation_evol_Looks_500_CDABr}
%%\end{figure}
%%%\begin{figure}[h!]
%%%\centering
%%%\includegraphics[width=0.75\figwidth]{images/Synth_estimation_DSP_SWH_500_Laplace.png}
%%%\caption{Estimated spectrum of SWH using the least square algorithm (red line)  and the proposed Brown-HMC algorithm (blue line).} \label{fig:Synth_estimation_DSP_SWH_500_Laplace}
%%%\end{figure}
\begin{table}[h] \centering
\centering \caption{Performance on synthetic BM data ($500$ echoes).}
\begin{tabular}{|c|c|c|c|c|c|c|}
  \cline{3-7}
\multicolumn{2}{c|}{}                        & SWH    & $\tau$  & \multirow{2}{*}{$P_u$}  & $\mu$ & \multirow{2}{*}{ENL} \\
\multicolumn{2}{c|}{}                        &  (cm)   &  (cm) &  & $\times 10^{(-4)}$ & \\
\hline  \multirow{3}{*}{Bias}   & LS-BM   & 3.18      & 1.1     & \textbf{-0.01} & $0.51$ & -\\
\cline{2-7}                     & CD-BM  & 0.32      & \textbf{0.08}    & -0.2  & $0.26$ & 0.97\\
\cline{2-7}                     & HMC-BM  & \textbf{-0.02}     & -0.13   & -0.2  & $-0.1$ & 0.93\\
\hline
\hline \multirow{3}{*}{STD}     & LS-BM   & 44.7    & 6.1   & 1.91 & $11$ & -\\
\cline{2-7}                     & CD-BM  & \textbf{2.72}   & \textbf{1.1}   & 0.62 & $12$  & 4.47 \\
\cline{2-7}                     & HMC-BM  & 5.74    & 1.8   & \textbf{0.52}  & $6$ & 4.46\\
\hline
\hline \multirow{1}{*}{Average time}& LS-BM   & \multicolumn{5}{c|}{8.9}  \\
\cline{2-7} \multirow{1}{*}{per echo }  & CD-BM  & \multicolumn{5}{c|}{\textbf{3.6}}  \\
\cline{2-7} \multirow{1}{*}{in milliseconds} & HMC-BM  & \multicolumn{5}{c|}{739.7}  \\
  \hline
\end{tabular}
\label{tab:Bias_STD_Synth}
\end{table}
%%%%%%%%%%%%%%%%%%%%%%%%%%%%%%%%%%%%%%%%%%%%%%%%%%%
%%%%%%%%%%%%%%%%%%%%%%%%%%%%%%%%%%%%%%%%%%%%%%%%%%%
%%%%%%%%%%%%%%%%%%%%%%%%%%%%%%%%%%%%%%%%%%%%%%%%%%%
%\clearpage
\section{Results on real data}
\label{sec:Results_on_real_data}
In order to evaluate the performance of the proposed algorithm on real data, we first present our comparison criteria. Then, two kinds of real data are considered: Jason-$2$ echoes for which BM is applied, and Cryosat-$2$ for which CA and DDA models are more appropriate.
%This section first introduces the evaluation criteria in the case of real data. Second, it evaluates the performance of the proposed algorithm for real Jason-$2$ echoes when considering BM. Third, it studies the Cryosat-$2$ CA and DDA echoes and compares the obtained results to those of the LS approach.

%%%%%%%%%%%%%%%%%%%%%%%%%%%%%%%%%%%%%%%%%%%%%%%%%%%
\subsection{Evaluation criteria} \label{subsec:Evaluation_criteria_real}
When processing real data, it is not possible to compare the performance of the different parameter estimation algorithms using the bias and STD defined in \eqref{eqt:bias} and \eqref{eqt:STD}, since the true parameter values are not known. In the case of real waveforms, it is usual to compare the different parameter estimations along a pass by plotting the different obtained results \cite{Sandwell2005,Amarouche2004}  and by considering a modified STD. The modified STD is computed using \eqref{eqt:STD} in which the true parameter value is approached by the mean of the estimated parameters along $20$ successive echoes. This modified STD is called ``STD at 20 Hz'' \cite{HalimiTGRS2014,Giles2012,Gommenginger2011OSTST}.

%In the case of  real waveforms, the mean of the estimated parameters is evaluated to compare the results of the different studied algorithms. The STD of the parameter is also evaluated. However,  the true parameters are not known for real data. This problem is dealt with in the altimetric community by approaching the true parameters of each successive $20$ echoes by their average (the number of echoes in $1$ second is $20$ echoes for Jason-$2$) \cite{HalimiTGRS2014}. The STDs can then be evaluated using \eqref{eqt:STD} and the approached true parameters. This criteria is known in the altimetric community as STD at $20$ Hz  \cite{HalimiTGRS2014}.

%%%%%%%%%%%%%%%%%%%%%%%%%%%%%%%%%%%%%%%%%%%%%%%%%%%
\subsection{Analysis of Jason-$2$ data with the Brown model}
\label{subsec:Conventional_altimetry_with_the_Brown_model}
\begin{figure}[h!]
\centering \subfigure[SWH]{\includegraphics[width=0.75\figwidth,height=6cm]{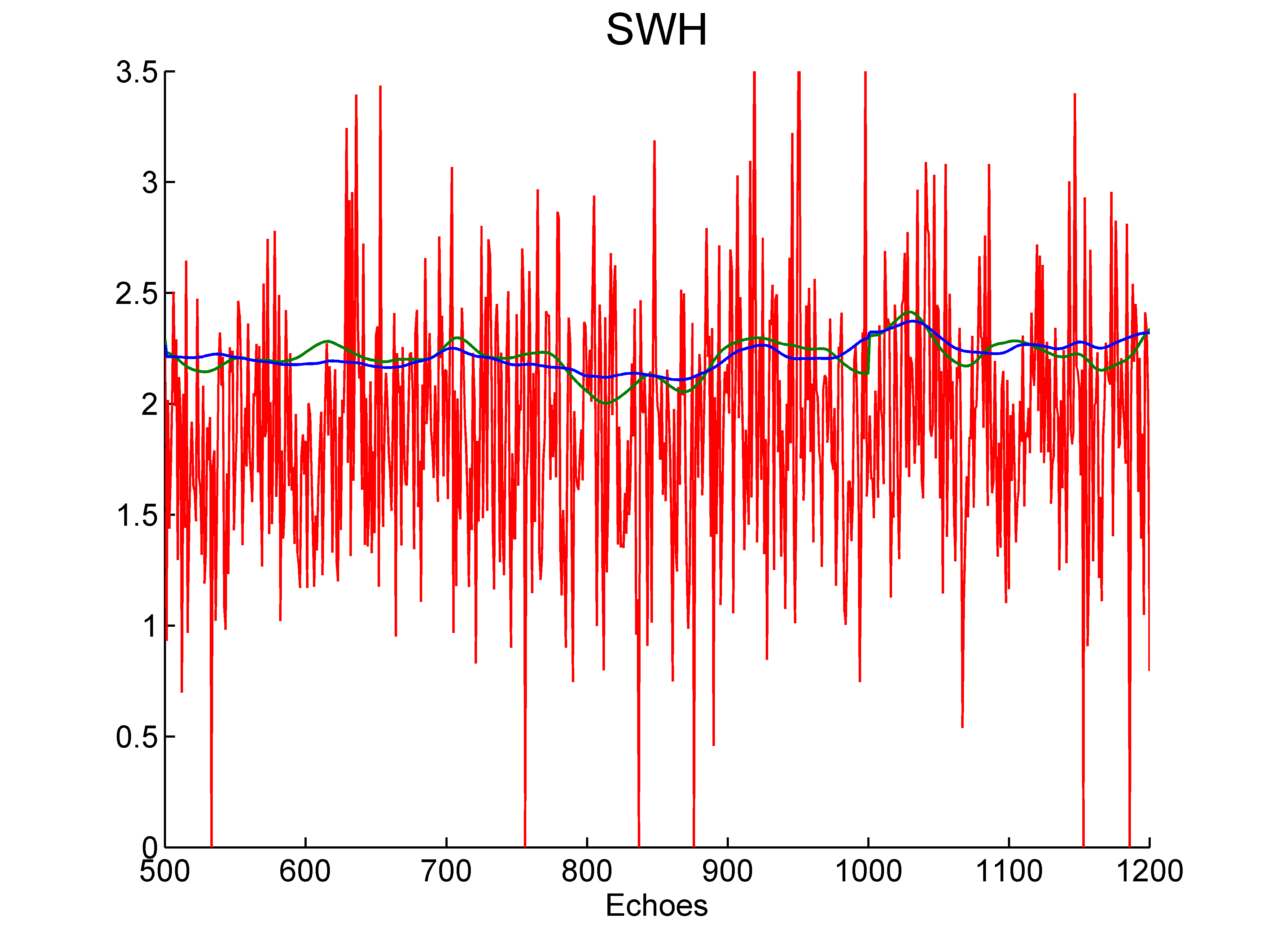}}
\subfigure[$\tau$]{\includegraphics[width=0.75\figwidth,height=6cm]{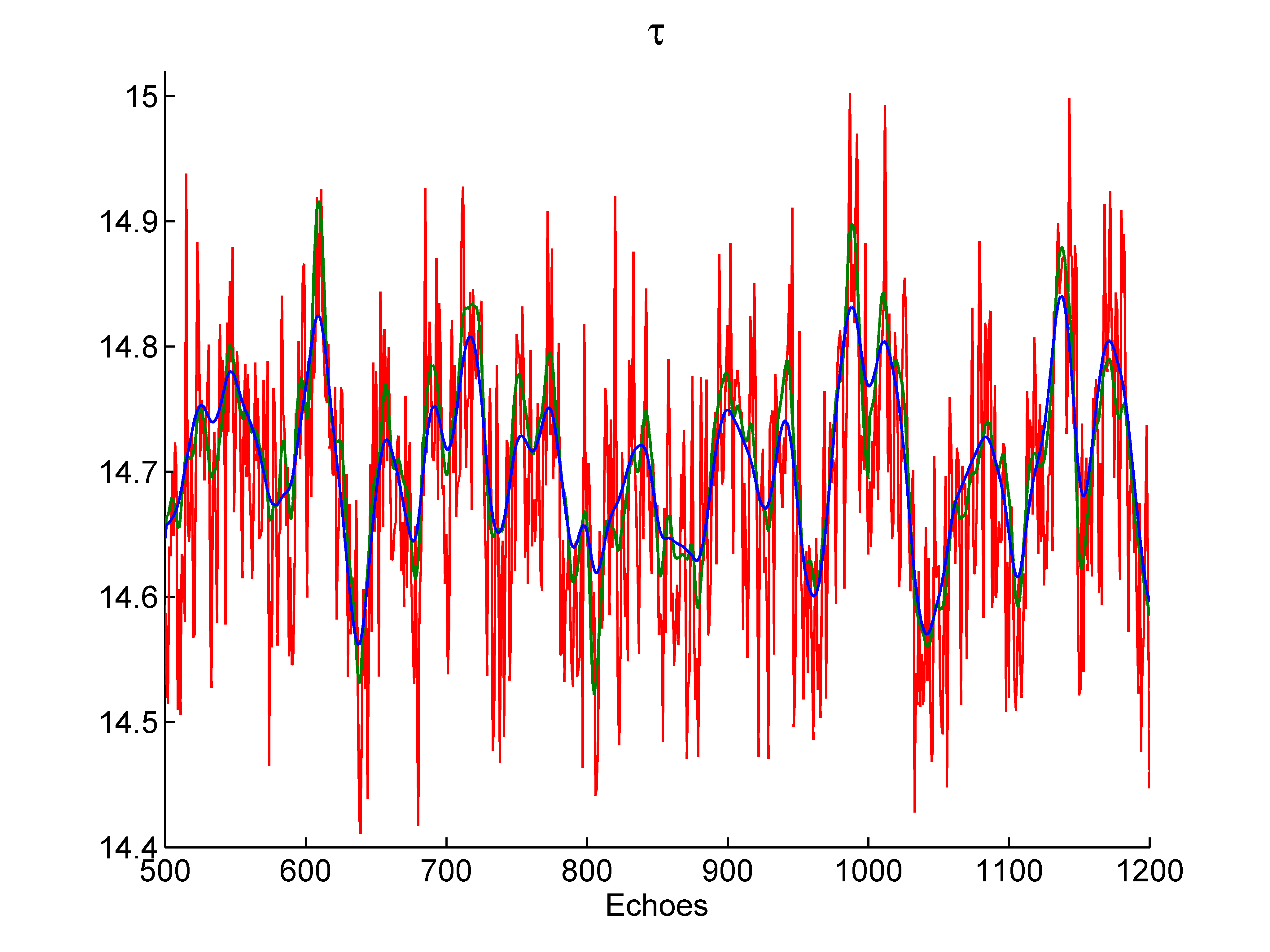}}
\subfigure[$P_u$]{\includegraphics[width=0.75\figwidth,height=6cm]{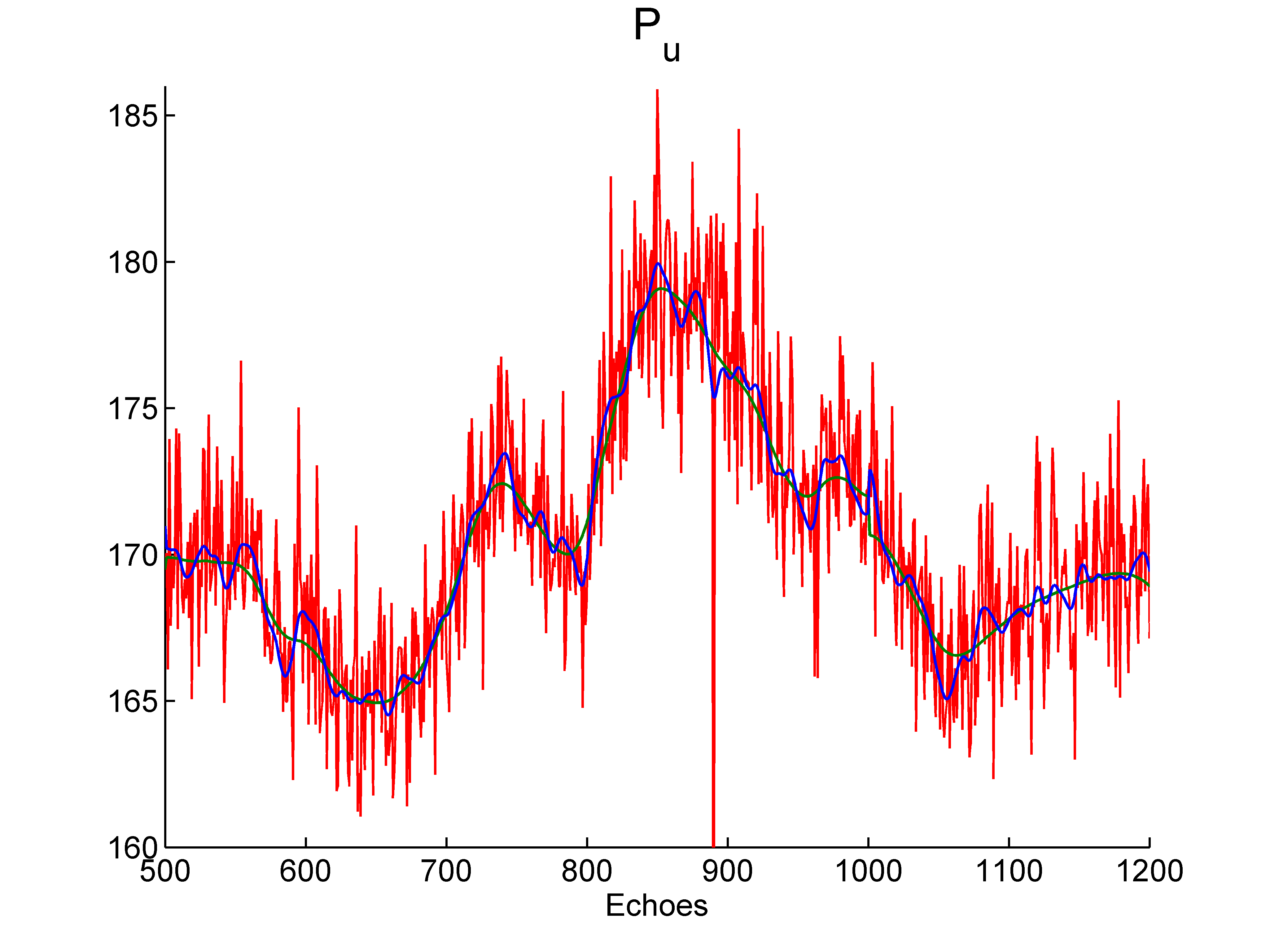}}
\caption{Estimated parameters using the LS-BM algorithm (red line), the HMC-BM algorithm (green line) and the CD-BM algorithm (blue line) for Jason-$2$ echoes. (a) SWH, (b) $\tau$ and (c) $P_u$. } \label{fig:Real_evol_Param_Jason2_CDABM_HMCBM}
\end{figure}

\begin{figure}[h!]
\centering
\includegraphics[width=0.75\figwidth]{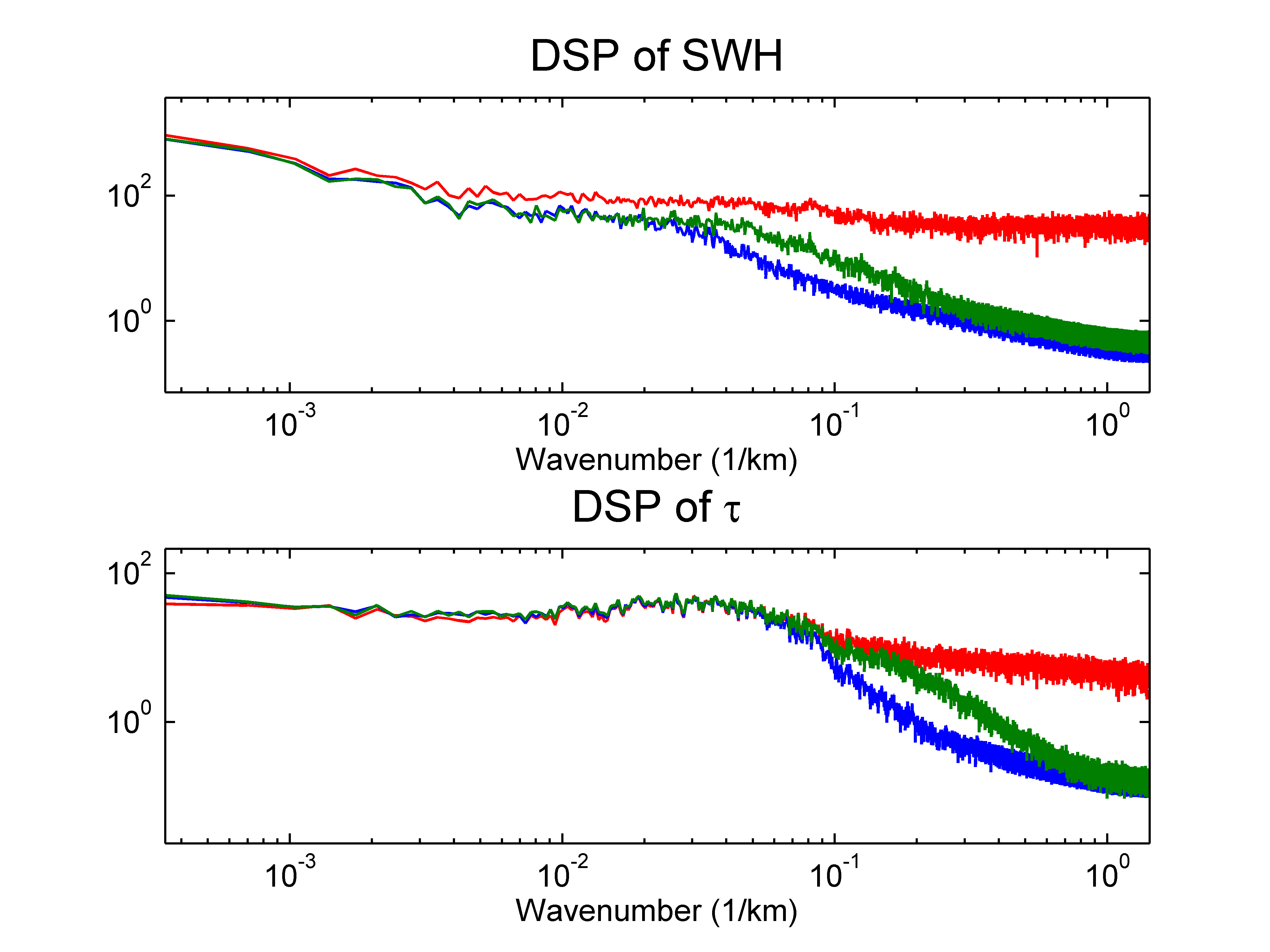}
\caption{Estimated spectra of SWH and $\tau$ using $43000$ Jason-2 echoes with the LS algorithm (red line), the HMC-BM algorithm (green line) and the CD-BM algorithm (blue line).} \label{fig:Real_DSPs_SWH_Tau_Jason2_CDABM_HMCBM}
\end{figure}

This section illustrates the performance of the proposed CD algorithm when applied to a real Jason-$2$ dataset. The considered data last $36$ minutes and consist  of $43000$ real echoes that were extracted  from the pass $30$ of cycle $35$. Figure \ref{fig:Real_evol_Param_Jason2_CDABM_HMCBM}   shows the parameters estimated on $700$ successive echoes along the pass when considering the  LS-BM (in red), HMC-BM (in green) and CD-BM (in blue) algorithms. As observed for synthetic data in Section \ref{sec:Simulations}, the LS-BM estimates present a high estimation noise while both CD and HMC algorithms provide  smoother estimates which are physically more consistent. Moreover, CD and HMC algorithms appear to be more robust to outliers as illustrated for the estimate $\# 890$ of $P_u$. Note that the estimated SWH is slightly larger for the Bayesian algorithms when compared to LS.  This difference can be explained by the i.i.d. noise assumption used in LS algorithm which is not in adequation with the considered data as already discussed in \cite{Sandwell2005,Halimi2013Eusipco}. The reduction of estimation variance provided by the Bayesian algorithms is particularly noticeable when plotting  the spectral power density of  SWH and $\tau$ as in Fig. \ref{fig:Real_DSPs_SWH_Tau_Jason2_CDABM_HMCBM}. This figure shows a clear noise reduction especially for ocean scales smaller than $100$ km for SWH and $10$ km for $\tau$ (note that we have not applied atmospheric and instrumental corrections on the estimated epoch $\tau$) which improves the observation of these scales. More quantitative results are provided in Table \ref{tab:Bias_STD_Time_Real}. This table shows a good agreement between the means of the estimated parameters for the LS, HMC and CD algorithms (except for SWH as explained above). Bayesian algorithms also estimate the ENL which provides an indication about the noise level corrupting the data. The estimated ENL ranges between $70$ and $80$ echoes which is close to the actual value (around $90$ or $100$ echoes). As for synthetic data, the estimated STDs obtained with the CD algorithm are smaller than for LS and HMC which is of great importance for many practical applications related to oceanography such as bathymetry. Indeed, the CD algorithm improves the LS-BM STDs by a factor of about $20$ for SWH, $2$ for $\tau$ and $3$ for $P_u$. Note finally that the proposed CD algorithm requires a reduced computational time  two times lower than that of the LS algorithm. As previously explained, this time  reduction is due to the parallel scheme considered by the CD algorithm which is more efficient than the independent processing of echoes performed in the LS algorithm.

%\begin{figure}[h!]
%\centering
%\includegraphics[width=0.7\figwidth]{images/Real_evol_SWH_Jason2_CDABM_HMCBM.png}
%\caption{Estimated SWH parameter using the LS algorithm (red line), the HMC-BM algorithm (green line) and the CD-BM algorithm (Blue line) for Jason-$2$ echoes.} \label{fig:Real_evol_SWH_Jason2_CDABM_HMCBM}
%\end{figure}
%\begin{figure}[h!]
%\centering
%\includegraphics[width=0.7\figwidth]{images/Real_evol_TAU_Jason2_CDABM_HMCBM.png}
%\caption{Estimated $\tau$ parameter using the LS algorithm (red line), the HMC-BM algorithm (green line) and the CD-BM algorithm (Blue line) for Jason-$2$ echoes.} \label{fig:Real_evol_TAU_Jason2_CDABM_HMCBM}
%\end{figure}
%\begin{figure}[h!]
%\centering
%\includegraphics[width=0.75\figwidth]{images/Real_evol_PU_Jason2_CDABM_HMCBM.png}
%\caption{Estimated $P_u$ parameter using the LS algorithm (red line), the HMC-BM algorithm (green line) and the CD-BM algorithm (Blue line) for Jason-$2$ echoes.} \label{fig:Real_evol_PU_Jason2_CDABM_HMCBM}
%\end{figure}
%\begin{figure}[h!]
%\centering
%\includegraphics[width=0.75\figwidth]{images/Real_estimation_evol_Nt_500_Laplace.png}
%\caption{Estimated $N_t$ parameter using the proposed Brown-HMC algorithm.} \label{fig:Real_estimation_evol_Nt_500_Laplace}
%\end{figure}
%\begin{figure}[h!]
%\centering
%\includegraphics[width=0.75\figwidth]{images/Real_estimation_evol_Looks_500_Laplace.png}
%\caption{Estimated number of looks (L) using the proposed Brown-HMC algorithm.} \label{fig:Real_estimation_evol_Looks_500_Laplace}
%\end{figure}
\begin{table}[h] \centering
\centering \caption{Performance on real Jason-$2$ data ($43000$ echoes).}
\begin{tabular}{|c|c|c|c|c|c|c|}
  \cline{3-7}
\multicolumn{2}{c|}{}                        & SWH (cm)   & $\tau$ (cm) & $P_u$ & $\mu$ & ENL\\
\hline  \multirow{3}{*}{Mean}   & LS-BM   & 237      & 14.67     & 167.81 & 0.97 &  -\\
\cline{2-7}                     & CD-BM  & 272      & 14.69     & 167.52 & 1.06 &  74\\
\cline{2-7}                     & HMC-BM  & 270      & 14.69     & 167.53 & 1.06 &  80\\
\hline
\hline \multirow{3}{*}{STD}     & LS-BM   & 55.5    & 11.23   & 8.18  & 0.12 & -\\
\cline{2-7}                     & CD-BM  & \textbf{2.37}   & \textbf{5.57}   & \textbf{2.79}  & 0.08 & 15\\
\cline{2-7}                     & HMC-BM  & 5.58    & 7.11   &  7.64 & 0.08 & 9.5\\
\hline
\hline \multirow{1}{*}{Average time}& LS-BM  & \multicolumn{5}{c|}{12.5}  \\
\cline{2-7} \multirow{1}{*}{per echo }  & CD-BM  & \multicolumn{5}{c|}{\textbf{6.6}}  \\
\cline{2-7} \multirow{1}{*}{in milliseconds} & HMC-BM  & \multicolumn{5}{c|}{1142.3}  \\
  \hline
\end{tabular}
\label{tab:Bias_STD_Time_Real}
\end{table}

%%%%%%%%%%%%%%%%%%%%%%%%%%%%%%%%%%%%%%%%%%%%%%%%%%%
\subsection{Comparison between CA and DDA for Cryosat-$2$ data}
\label{subsec:Comparison_between_CA_and_DDA_for_Cryosat_2_data}
\begin{figure}[h!]
\centering
\includegraphics[width=0.75\figwidth]{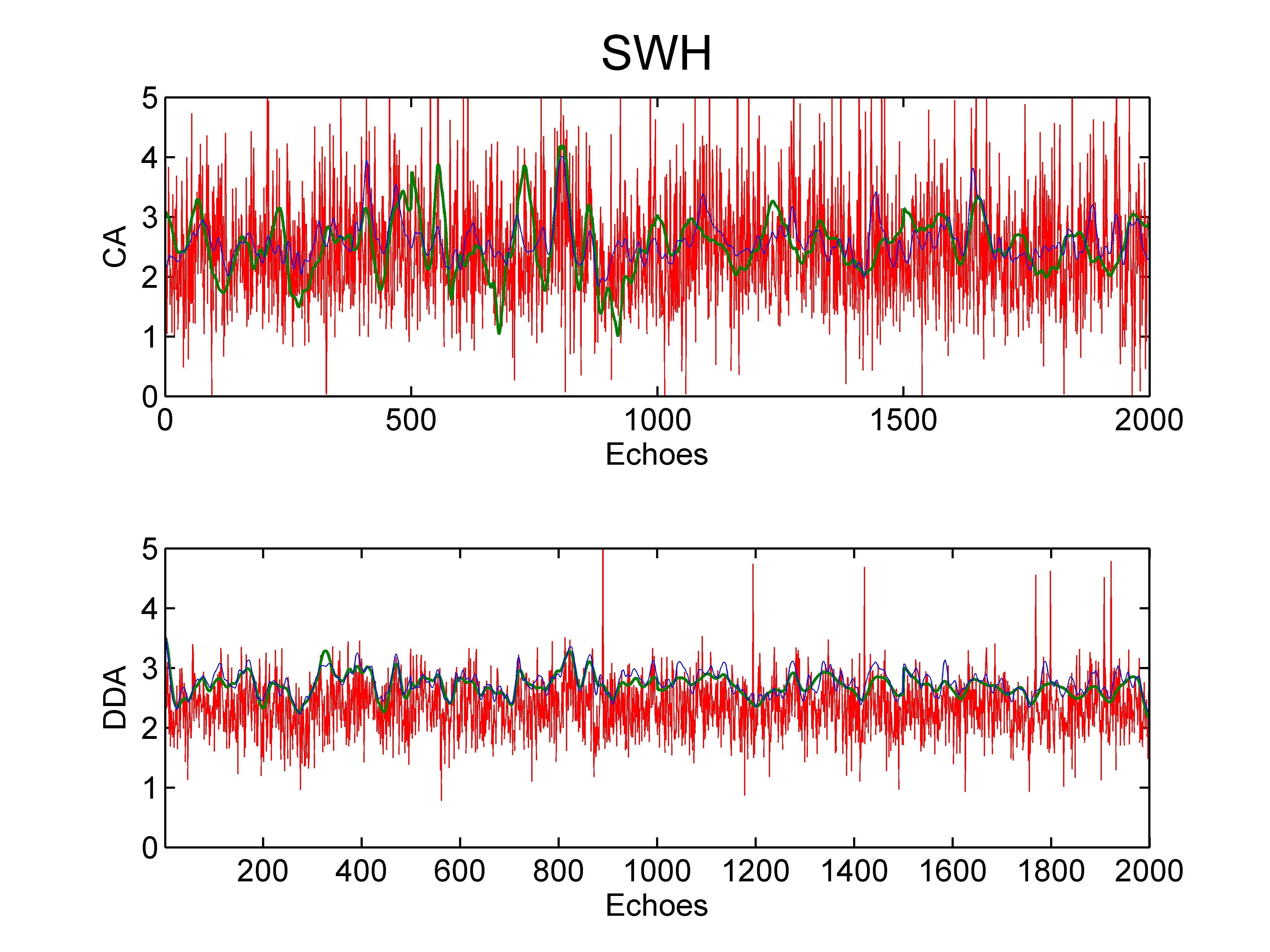}
\caption{Estimated SWH using $2000$ Cryosat-$2$ echoes with the LS algorithm (red line), the HMC algorithm (green line) and the CD algorithm (blue line). (Top) CA echoes, (bottom) DDA echoes.} \label{fig:Real_evol_SWH_Cryosat2_CA_DDA}
\end{figure}

\begin{figure}[h!]
\centering
\includegraphics[width=0.75\figwidth]{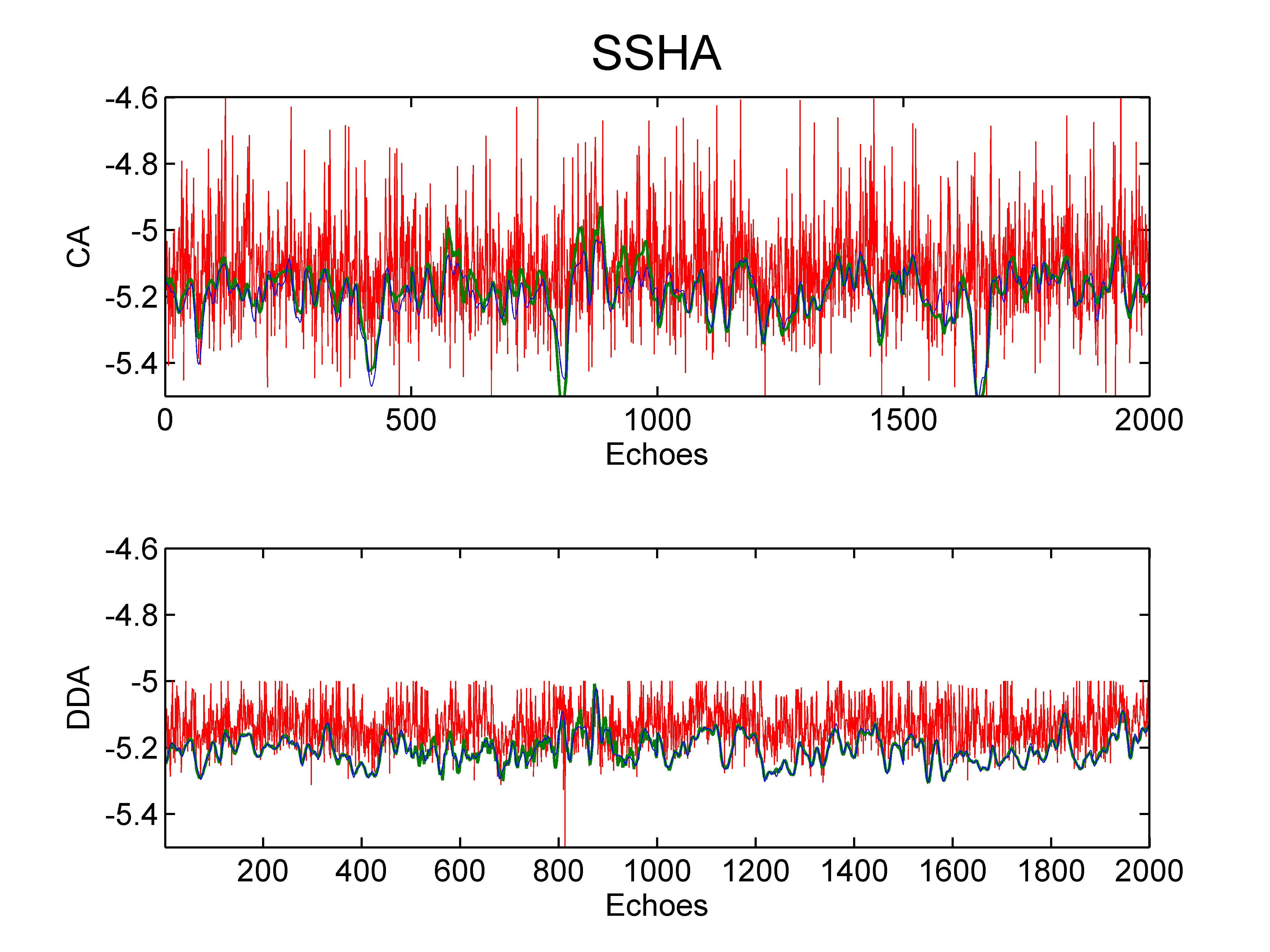}
\caption{Estimated SSHA using $2000$ Cryosat-$2$ echoes with the LS algorithm (red line), the HMC algorithm (green line) and the CD algorithm (blue line). (Top) CA echoes, (bottom) DDA echoes.} \label{fig:Real_evol_TAU_Cryosat2_CA_DDA}
\end{figure}

This section evaluates the performance of the proposed CD  algorithm for real oceanic Cryosat-$2$ waveforms. The Cryosat-$2$ altimeter called SIRAL presents three modes that are: the low resolution mode (LRM), the synthetic aperture radar mode (SARM) and the synthetic aperture radar interferometric mode (SARInM) \cite{Wingham2006}. This paper considers data from the SARM that are used to generate CA\footnote{These echoes are known under different names: LRM-like \cite{Phalippou2011}, pseudo-LRM
\cite{Smith2012,Gommenginger2012OSTST,Boy2012},
reduced-SAR (RDSAR) \cite{Gommenginger2012OSTST,Boy2012} or CA-SARM in \cite{HalimiTGRS2014}. These echoes are used to compare the parameters of CA with those of DDA.}  and DDA echoes (see \cite{HalimiTGRS2014} for more details about these echoes).
The considered dataset lasts approximately $100$ seconds (i.e., $2000$ echoes) and was obtained in August $2011$ by the Cryosat processing prototype (CPP) developed by CNES \cite{Boy2012} (which is doing the level $1$ processing). Note that the CNES-CPP uses data pre-processed to full bit rate provided by ESA \cite{Wingham2006}. Table \ref{tab:Bias_STD_Real} compares the estimated parameters using the LS algorithm  with the CD  and HMC algorithms when considering both CA and DDA echoes. For all algorithms, we have considered the same instrumental parameters as in \cite{HalimiTGRS2014,HalimiTGRS2014b} which are given in Table \ref{tab:valeurs_satellite}. To compare with \cite{HalimiTGRS2014}, we only present the estimation results of SWH and of the sea surface height anomaly (SSHA)  which represent the most important parameters ($\textrm{SSHA}$  is obtained by applying all environmental corrections on the estimated epoch $\tau$). Figs. \ref{fig:Real_evol_SWH_Cryosat2_CA_DDA} and \ref{fig:Real_evol_TAU_Cryosat2_CA_DDA}
show the evolution of the estimated SWH and SSHA parameters when considering the three studied algorithms. These figures show a good agreement between the estimated parameters especially between CD and HMC algorithms, which is confirmed by the mean of the estimated parameters provided in Table \ref{tab:Bias_STD_Real}.  Note that, as for Jason-$2$, a bias seems to appear between the estimated SWH parameter for Bayesian algorithms and the LS approach, which can be explained similarly. Considering the STDs, and to compare the results of Table \ref{tab:Bias_STD_Real} with \cite{HalimiTGRS2014}, one has to note that the results of \cite{HalimiTGRS2014} were obtained when using the LS algorithm for both CA and DDA. The processing of the Cryosat-$2$ data further highlights the main advantage of the proposed CD approach which is to reduce drastically the STDs of the estimated parameters. For example,  STDs(SWH) (resp. STD(SSHA))  goes from $91$ cm (resp. $14$ cm) for LS-CA to $3$ cm (resp. $4$ cm) for CD-CA which represents a great improvement. The same behavior is observed when considering DDA echoes showing the interest of smoothing the altimetric parameters. The second advantage of the CD algorithm is the reduction of the computational time especially for DDA echoes. Indeed,  Table \ref{tab:Bias_STD_Real} highlights a time reduction by a factor $20$ w.r.t. LS-DDA that processed the echoes independently while CD-DDA uses a parallel scheme. For CA echoes, the CD approach shows a time  improvement by a factor $1.5$ which is still an interesting result. Note finally that the HMC algorithm shows similar performance than the CD algorithm but at a price of a higher computation cost. These results confirm the good performance of the proposed algorithm and show its flexibility w.r.t. the considered altimetric models.
\begin{table}[h!]
\centering
\begin{tabular}{|c|l|}
\hline
          Parameter                               & Value \\
  \hline  Frequency                               & $13.575$ GHz\\
          Wavelength ($\lambda$)                  & $2.21$ cm\\
          Bandwidth (B)                           & $320$ MHz \\
          Altitude ($h$)                          & $730$ km \\
          Burst repetition frequency (BRF)        & $85$ Hz \\
          Pulse repetition frequency (PRF)        & $18182$ Hz \\
          3 dB Antenna beam width ($\theta_{3\textrm{dB}}$)& $1.1388$ degrees \\
          Velocity ($v_{s}$)                      & $7000$ m/s \\
          Pulses per burst                        & $64$ pulses \\
          Burst length ($\tau_b$)                 & $3.5$ ms \\
          Doppler beam width                      & $327$ m
          \\
  \hline
\end{tabular}
\caption{Instrumental  parameters.}
\label{tab:valeurs_satellite}
\end{table}
\begin{table}[h] \centering
\centering \caption{Comparison between the LS, CD and HMC algorithms when considering real CA and DDA data $\;\;$ ($2000$ Cryosat-$2$ echoes).}
\begin{tabular}{|c|c|c|c|c||c|c|c|}
  \cline{3-8}
\multicolumn{2}{c|}{}    & \multicolumn{3}{c||}{CA echoes}  & \multicolumn{3}{c|}{DDA echoes} \\
  \cline{3-8}
\multicolumn{2}{c|}{}                         & LS-CA  & CD-CA & HMC-CA  & LS-DDA  & CD-DDA   & HMC-DDA \\
\hline  \multirow{2}{*}{Means}   & SWH (m)  &  2.39 &  2.59 &  2.53 &  2.31 &  2.70 &  2.70 \\
\cline{2-8}                      & SSHA (m) & -5.12 & -5.20 & -5.19 & -5.13 & -5.20 & -5.21 \\

\hline  \multirow{2}{*}{STDs at 20 Hz} & SWH (cm) &  91 &  \textbf{3} &  20 &  45 &  \textbf{3} &  8 \\
\cline{2-8}                           & SSHA (cm) &  14 &  \textbf{4} &  \textbf{4} &  6 &  \textbf{2} &  \textbf{2} \\
\hline  \multicolumn{2}{|c|}{Average time }     & \multirow{2}{*}{0.1} & \multirow{2}{*}{\textbf{0.07}} & \multirow{2}{*}{252} & \multirow{2}{*}{0.79}  & \multirow{2}{*}{\textbf{0.04}}  & \multirow{2}{*}{233} \\
  \multicolumn{2}{|c|}{per echo (s)}      &   &   &   &   &  &   \\
  \hline
\end{tabular}
\label{tab:Bias_STD_Real}
\end{table}
%%%%%%%%%%%%%%%%%%%%%%%%%%%%%%%%%%%%%%%%%%%%%%%%%%%
%%%%%%%%%%%%%%%%%%%%%%%%%%%%%%%%%%%%%%%%%%%%%%%%%%%
%%%%%%%%%%%%%%%%%%%%%%%%%%%%%%%%%%%%%%%%%%%%%%%%%%%
\clearpage
\section{Conclusions} \label{sec:Conclusions}
This paper proposed a new Bayesian strategy for a smooth  parameter estimation of the noise  and altimetric parameters. The altimetric parameters were assigned a prior distribution enforcing a smooth temporal evolution. The altimetric signal was assumed to be corrupted by a thermal and speckle noise distributed according to an independent and non identically  Gaussian distribution. The parameter estimation was achieved using a coordinate descent algorithm that presents a low computational cost which is suitable for real time applications. The proposed CD  algorithm showed good performance and improved the quality of the estimated parameters when applied to both synthetic and real altimetric echoes from conventional and delay/Doppler altimetry. More precisely, it provided reduced parameter STDs  and presented a lower computational cost when compared to the state-of-the-art LS algorithm. Moreover, the CD  algorithm is generic and showed its efficiency when considering three altimetric models. Future work includes the consideration of other altimetric models such as the DDA-SAMOSA model \cite{Puig2008,Gommenginger2012}, coastal models \cite{Gomez2009,HalimiTGRS2012} and the CA model based on the true PTR \cite{AmaroucheOSTST2014,BoyOSTST2014}. Considering antenna mispointing is also an interesting issue that is currently under investigation.

\section*{Acknowledgments}
The authors would like to thank CLS, CNES and ESA for providing real altimetric data.

%%%%%%%%%%%%%%%%%%%%%%%%%%%%%%%%%%%%%%%%%%%%%%%%%%%
%%%%%%%%%%%%%%%%%%%%%%%%%%%%%%%%%%%%%%%%%%%%%%%%%%%
%%%%%%%%%%%%%%%%%%%%%%%%%%%%%%%%%%%%%%%%%%%%%%%%%%%
\clearpage
\appendix[Mathematical derivations] \label{app:Mathematical_derivations}

%%%%%%%%%%%%%%%%%%%%%%%%%%%%%%%%%%%%%%%%%%%%%%%%%%%
\subsection{Hybrid Gibbs algorithm} \label{app:Hybrid_Gibbs_algorithm}
The principle of the Gibbs sampler is to sample according to the conditional distributions of the posterior of interest \cite{Robert2007}. The MCMC based algorithm proposed in \cite{Halimi2015ICASSP} uses this principle to sequentially sample the parameters $\bThe, \bmu, \bLam$ and $\beps$. When a conditional distribution cannot be sampled directly, sampling techniques such as the HMC algorithm can be applied (see \cite{Brooks2011,Halimi_ArXiv2014} for more details about this algorithm). This algorithm has shown better mixing property than independent or random walk Metropolis-Hasting moves especially in high-dimensional problems \cite{Brooks2011,Halimi_ArXiv2014}. Therefore, it has been considered in \cite{Halimi2015ICASSP}  since the vectors to sample are of size ($M \times 1$). The Gibbs sampler (including the HMC procedures) used to sample according to the posterior \eqref{eqt:Joint_Posterior} consists of the four steps summarized in Algo. \ref{alg:Hybrid_Gibbs_sampler}.
\begin{algorithm}
\caption{Hybrid Gibbs sampler} \label{alg:Hybrid_Gibbs_sampler}
\begin{algorithmic}[1]
       \STATE \underline{Initialization}
       \STATE Initialize parameters $\bthe_{i}^{(0)},$ for $i \in  \left\lbrace 1,\cdots,3 \right\rbrace$, $\bmu^{(0)}$, $\bLam^{(0)}$ and $\beps^{(0)}$
       \STATE \underline{Update parameters/hyperparameters}
       \FOR{$t=1:N_{\textrm{bi}}+N_{r}$}
               \STATE Sample $\bthe_{i}^{(t)},$ for $i \in  \left\lbrace 1,2,3 \right\rbrace$ according to \eqref{eqt:posterior_Theta_i} using HMC
               %\STATE Sample $\bthe_{3}^{(t)},$  according to \eqref{eqt:posterior_Pu}
               \STATE Sample $\bmu^{(t)}$ according to \eqref{eqt:posterior_mu_m}
               \STATE Sample $\bLam^{(t)}$ according to \eqref{eqt:posterior_sigma_nk}
               \STATE Sample $\beps^{(t)}$ according to \eqref{eqt:posterior_epsilon_i}
       \ENDFOR
\end{algorithmic}
\end{algorithm}
This algorithm uses the conditional distribution associated with the altimetric parameters and the hyperparameters that are obtained by considering the likelihood \eqref{eqt:likelihood_globale} and the priors \eqref{eqt:prior_parameters} and \eqref{eqt:prior_epsilon}, leading to
\begin{equation}
f(\bthe_{i}|\bsY,\bThe_{\backslash i}, \bmu, \bLam,\epsilon^2_i) \propto \exp{\left(-\frac{1}{2 \epsilon_i^2}  \left\| \bsD \bthe_{i}\right\|^2 \right)}  \exp
\left\lbrace - \sum_{m=1}^{M}{ \frac{\bsx_m^T  \bSig_m^{-1} \bsx_m}{2}  }  \right\rbrace
\label{eqt:posterior_Theta_i}
\end{equation}
and
\begin{equation}
\epsilon_i^2 | \bsY, \bThe, \bmu, \bSig  \sim   \calI \calG \left(\frac{M}{2}+a_i, \frac{\left\| \bsD \bthe_{i}\right\|^2}{2} + b_i\right)
\label{eqt:posterior_epsilon_i}
\end{equation}
for $i \in \left\lbrace 1,\cdots,3\right\rbrace$.
Note that Algo. \ref{alg:Hybrid_Gibbs_sampler} draws $N_{\textrm{MC}} = N_{\textrm{bi}} + N_{r}$ samples that are used to approximate the MMSE estimator of the parameters after removing the first $N_{\textrm{bi}}$ burn-in samples. This algorithm benefits from the robust properties of the MCMC methods and allows the estimation of the hyperparameters. However, it presents a high computational cost that prevents its use for real time applications.

%\newpage
\bibliographystyle{ieeetran} 
\bibliography{biblio}

\begin{thebibliography}{10}
\providecommand{\url}[1]{#1}
\def\UrlFont{\rmfamily}
\providecommand{\newblock}{\relax}
\providecommand{\bibinfo}[2]{#2}
\providecommand\BIBentrySTDinterwordspacing{\spaceskip=0pt\relax}
\providecommand\BIBentryALTinterwordstretchfactor{4}
\providecommand\BIBentryALTinterwordspacing{\spaceskip=\fontdimen2\font plus
\BIBentryALTinterwordstretchfactor\fontdimen3\font minus
  \fontdimen4\font\relax}
\providecommand\BIBforeignlanguage[2]{{%
\expandafter\ifx\csname l@#1\endcsname\relax
\typeout{** WARNING: IEEEtran.bst: No hyphenation pattern has been}%
\typeout{** loaded for the language `#1'. Using the pattern for}%
\typeout{** the default language instead.}%
\else
\language=\csname l@#1\endcsname
\fi
#2}}

\bibitem{Vincent2006}
P.~Vincent, N.~Steunou, E.~Caubet, L.~Phalippou, L.~Rey, E.~Thouvenot, and
  J.~Verron, ``{AltiKa: a Ka-band altimetry payload and system for operational
  altimetry during the GMES period},'' \emph{Sensors}, vol.~6, no.~3, pp.
  208--234, Mar. 2006.

\bibitem{Richard2007}
J.~Richard, L.~Phalippou, F.~Robert, N.~Stenou, E.~Thouvenot, and P.~Sengenes,
  ``An advanced concept of radar altimetry over oceans with improved
  performances and ocean sampling : {AltiKa},'' in \emph{Proc. IEEE Int. Conf.
  Geosci. Remote Sens. (IGARSS)}, July 2007, pp. 3537--3540.

\bibitem{Raney2011}
R.~K. Raney and L.~Phalippou, ``The future of coastal altimetry,'' in
  \emph{Coastal Altimetry}, S.~Vignudelli, A.~G. Kostianoy, P.~Cipollini, and
  J.~Benveniste, Eds.\hskip 1em plus 0.5em minus 0.4em\relax Berlin: Springer
  Verlag, 2011, pp. 61--101.

\bibitem{Raney1998}
R.~K. Raney, ``The delay/{D}oppler radar altimeter,'' \emph{IEEE Trans. Geosci.
  Remote Sens.}, vol.~36, no.~5, pp. 1578--1588, Sept. 1998.

\bibitem{Wingham2004}
D.~Wingham, L.~Phalippou, C.~Mavrocordatos, and D.~Wallis, ``The mean echo and
  echo cross product from a beamforming interferometric altimeter and their
  application to elevation measurement,'' \emph{IEEE Trans. Geosci. Remote
  Sens.}, vol.~42, no.~10, pp. 2305 -- 2323, Oct. 2004.

\bibitem{HalimiTGRS2014}
A.~Halimi, C.~Mailhes, J.-Y. Tourneret, P.~Thibaut, and F.~Boy, ``A
  semi-analytical model for delay/{D}oppler altimetry and its estimation
  algorithm,'' \emph{IEEE Trans. Geosci. Remote Sens.}, vol.~52, no.~7, pp.
  4248--4258, July 2014.

\bibitem{HalimiTGRS2014b}
A.~Halimi, C.~Mailhes, J.-Y. Tourneret, T.~Moreau, and F.~Boy, ``{I}ncluding
  antenna mispointing in a semi-analytical model for delay/{D}oppler
  altimetry,'' \emph{IEEE Trans. Geosci. Remote Sens.}, vol.~53, no.~2, pp.
  598--608, Feb. 2015.

\bibitem{Brown1977}
G.~Brown, ``The average impulse response of a rough surface and its
  applications,'' \emph{IEEE Trans. Antennas and Propagation}, vol.~25, no.~1,
  pp. 67--74, Jan. 1977.

\bibitem{Amarouche2004}
L.~Amarouche, P.~Thibaut, O.~Z. Zanife, J.-P. Dumont, P.~Vincent, and
  N.~Steunou, ``Improving the {J}ason-1 ground retracking to better account for
  attitude effects,'' \emph{Marine Geodesy}, vol.~27, no. 1-2, pp. 171--197,
  Aug. 2004.

\bibitem{HalimiTGRS2012}
A.~Halimi, C.~Mailhes, J.-Y. Tourneret, P.~Thibaut, and F.~Boy, ``Parameter
  estimation for peaky altimetric waveforms,'' \emph{IEEE Trans. Geosci. Remote
  Sens.}, vol.~51, no.~3, pp. 1568--1577, March 2013.

\bibitem{Halimi2013}
\BIBentryALTinterwordspacing
A.~Halimi, ``From conventional to delay/doppler altimetry,'' Ph.D.
  dissertation, Institut National Polytechnique de Toulouse, Toulouse, France,
  2013. [Online]. Available:
  \url{http://ethesis.inp-toulouse.fr/archive/00002423/01/halimi.pdf}
\BIBentrySTDinterwordspacing

\bibitem{Maus1998}
S.~Maus, C.~M. Green, and J.~D. Fairhead, ``Improved ocean-geoid resolution
  from retracked {ERS-1} satellite altimeter waveforms,'' vol. 134, no.~1, pp.
  243--253, Feb. 1998.

\bibitem{Sandwell2005}
D.~T. Sandwell and W.~H.~F. Smith, ``Retracking {ERS-1} altimeter waveforms for
  optimal gravity field recovery,'' \emph{Geophys. J. Int.}, vol. 163, no.~1,
  pp. 79--89, Oct. 2005.

\bibitem{Severini2008}
J.~Severini, C.~Mailhes, J.-Y. Tourneret, and P.~Thibaut, ``Bayesian estimation
  of altimeter echo parameters,'' in \emph{Proc. IEEE Int. Conf. Geosci. Remote
  Sens. (IGARSS)}, vol.~3, Boston, MA, 2008, pp. 238--241.

\bibitem{Dumont1985}
J.-P. Dumont, ``Estimation optimale des param\`etres altim\'etriques des
  signaux radar {P}os\'eidon,'' Ph.D. dissertation, Institut National
  Polytechnique de Toulouse, Toulouse, France, 1985.

\bibitem{Halimi2013Eusipco}
A.~Halimi, C.~Mailhes, and J.-Y. Tourneret, ``Cram\'er-{R}ao bounds and
  estimation algorithms for delay/{D}oppler and conventional altimetry,'' in
  \emph{Proc. EUSIPCO}, Marrakech-Marocco, Sept. 9-13 2013.

\bibitem{Puig2009}
C.~Martin-Puig and G.~Ruffini, ``{SAR} altimeter retracker performance bound
  over water surfaces,'' in \emph{Proc. IEEE Int. Conf. Geosci. Remote Sens.
  (IGARSS)}, Cape Town, South Africa, July 12-17, 2009, pp. 449 --452.

\bibitem{Campisi2007}
P.~Campisi and K.~Egiazarian, \emph{Blind image deconvolution: theory and
  applications}.\hskip 1em plus 0.5em minus 0.4em\relax Boca Raton, FL: Taylor
  $\&$ Francis, cop. 2007, 2007.

\bibitem{Molina2006}
R.~Molina, J.~Mateos, and A.~Katsaggelos, ``Blind deconvolution using a
  variational approach to parameter, image, and blur estimation,'' \emph{IEEE
  Trans. Image Process.}, vol.~15, no.~12, pp. 3715--3727, Dec. 2006.

\bibitem{Sigurdsson2014}
J.~Sigurdsson, M.~Ulfarsson, and J.~Sveinsson, ``Hyperspectral unmixing with
  $l_{q}$ regularization,'' \emph{IEEE Trans. Geosci. Remote Sens.}, vol.~52,
  no.~11, pp. 6793--6806, Nov. 2014.

\bibitem{Chaari2013}
L.~Chaari, T.~Vincent, F.~Forbes, M.~Dojat, and P.~Ciuciu, ``Fast joint
  detection-estimation of evoked brain activity in event-related fmri using a
  variational approach,'' \emph{IEEE Trans. Image Processing}, vol.~32, no.~5,
  pp. 821--837, May 2013.

\bibitem{Halimi2015ICASSP}
A.~Halimi, C.~Mailhes, and J.-Y. Tourneret, ``Nonlinear regression using smooth
  {B}ayesian estimation,'' in \emph{Proc. IEEE Int. Conf. Acoust., Speech, and
  Signal Process. (ICASSP)}, Brisbane, Australia, April 2015, to appear.

\bibitem{Brooks2011}
S.~Brooks, A.~Gelman, G.~L.~. Jones, and X.-L. Meng, \emph{{Handbook of Markov
  chain Monte Carlo}}.\hskip 1em plus 0.5em minus 0.4em\relax ser. Chapman $\&$
  Hall/CRC Handbooks of Modern Statistical Methods. Taylor $\&$ Francis, 2011.

\bibitem{Bertsekas1995}
D.~P. Bertsekas, \emph{Nonlinear programming}.\hskip 1em plus 0.5em minus
  0.4em\relax Belmont, Massachusetts: Athena Scientific, 1995.

\bibitem{Amari1998}
S.~Amari and S.~Douglas, ``Why natural gradient?'' in \emph{Proc. IEEE
  ICASSP-98}, vol.~2, May 1998, pp. 1213--1216.

\bibitem{Green1984}
P.~J. Green, ``Iteratively reweighted least squares for maximum likelihood
  estimation, and some robust and resistant alternatives,'' \emph{Journal of
  the Royal Statistical Society. Series B (Methodological)}, vol.~46, no.~2,
  pp. 149--192, 1984.

\bibitem{Girolami2011}
M.~Girolami and B.~Calderhead., ``Riemann manifold {L}angevin and {H}amiltonian
  {M}onte {C}arlo methods,'' \emph{Journal of the Royal Statistical Society,
  Series B}, vol.~73, no.~2, pp. 123--214, March 2011.

\bibitem{Pereyra2013}
M.~Pereyra, H.~Batatia, and S.~McLaughlin, ``Exploiting information geometry to
  improve the convergence properties of variational active contours,''
  \emph{Selected Topics in Signal Process., IEEE Journal of}, vol.~7, no.~4,
  pp. 700--707, Aug. 2013.

\bibitem{AmaroucheOSTST2014}
L.~{A}marouche \emph{et al}, ``{Reduction of the Sea Surface Height spectral
  hump using a new retracker decorrelating ocean estimated parameters
  (DCORE)},'' in \emph{Report of the Ocean Surf. Topogr. Sci. Team Meet.
  (OSTST)}, Lake Constance, Germany, Oct. 27-31, 2014.

\bibitem{BoyOSTST2014}
F.~{B}oy \emph{et al}, ``{Towards the Jason-3 waveform processing: assessment
  of the numerical retracking performances},'' in \emph{Report of the Ocean
  Surf. Topogr. Sci. Team Meet. (OSTST)}, Lake Constance, Germany, Oct. 27-31,
  2014.

\bibitem{Hayne1980}
G.~Hayne, ``Radar altimeter mean return waveforms from near-normal-incidence
  ocean surface scattering,'' \emph{IEEE Trans. Antennas and Propagation},
  vol.~28, no.~5, pp. 687--692, Sept. 1980.

\bibitem{Phalippou2011}
\BIBentryALTinterwordspacing
L.~Phalippou and F.~Demeestere, ``Optimal re-tracking of {SAR} altimeter echoes
  over open ocean: from theory to results for {SIRAL2},'' in \emph{Report of
  the Ocean Surf. Topogr. Sci. Team Meet. (OSTST)}, San Diego, California, Oct.
  19-21, 2011. [Online]. Available:
  \url{http://www.aviso.oceanobs.com/fileadmin/documents/OSTST/2011/oral/01_Wednesday/Splinter%201%20IP/02%20OSTST-2011-Phalippou-Demeestere.pdf}
\BIBentrySTDinterwordspacing

\bibitem{Germain2006}
O.~Germain and G.~Ruffini, ``A revisit to the {GNSS-R} code range precision,''
  in \emph{Proc. GNSS-R}, Noordwijk, The Netherlands, June 14-15, 2006.

\bibitem{Puig2010}
C.~Martin-Puig, P.~Berry, R.~Smith, C.~Gommenginger, G.~Ruffini, P.~Cipollini,
  L.~Stenseng, A.~O., P.~Cotton, J.~Benveniste, and S.~Dinardo, ``{SAR}
  altimetry over water surfaces,'' in \emph{Oceans from Space}, Venice, Italy,
  April 2010.

\bibitem{Mazet2011}
V.~Mazet, ``Joint bayesian decomposition of a spectroscopic signal sequence,''
  \emph{IEEE Signal Processing Lett.}, vol.~18, no.~3, pp. 181--184, March
  2011.

\bibitem{Giles2012}
\BIBentryALTinterwordspacing
K.~Giles, D.~Wingham, N.~Galin, R.~Cullen, and W.~Smith, ``Precise estimates of
  ocean surface parameters from cryosat,'' in \emph{Report of the Ocean Surf.
  Topogr. Sci. Team Meet. (OSTST)}, Venice-Lido, Italy, Sept. 22-29, 2012.
  [Online]. Available:
  \url{http://www.aviso.oceanobs.com/fileadmin/documents/OSTST/2012/oral/02_friday_28/05_instr_processing_IIb/02_IP2B_Giles.pdf}
\BIBentrySTDinterwordspacing

\bibitem{Gommenginger2011OSTST}
\BIBentryALTinterwordspacing
C.~Gommenginger, C.~Martin-Puig, S.~Dinardo, D.~Cotton, M.~Srokosz, and
  J.~Benveniste, ``{I}mproved altimetric accuracy of {SAR} altimeters over the
  ocean: observational evidence from {C}ryosat-2 {SAR} and {J}ason-2,'' in
  \emph{Report of the Ocean Surf. Topogr. Sci. Team Meet. (OSTST)}, San Diego,
  California, Oct. 19-21, 2011. [Online]. Available:
  \url{http://www.aviso.oceanobs.com/fileadmin/documents/OSTST/2011/oral/01_Wednesday/Splinter%201%20IP/01%20Gommenginger%20SAMOSA_OSTST_Oct2011.pdf}
\BIBentrySTDinterwordspacing

\bibitem{Wingham2006}
D.~Wingham, C.~R. Francis, S.~Baker, C.~Bouzinac, D.~Brockley, R.~Cullen,
  P.~de~Chateau-Thierry, S.~Laxon, U.~Mallow, C.~Mavrocordatos, L.~Phalippou,
  G.~Ratier, L.~Rey, F.~Rostan, P.~Viau, and D.~Wallis, ``Cryosat: A mission to
  determine the fluctuations in {E}arth's land and marine ice fields,''
  vol.~37, no.~4, pp. 841 -- 871, July 2006.

\bibitem{Smith2012}
\BIBentryALTinterwordspacing
W.~H.~F. Smith and R.~Scharroo, ``Pulse-to-pulse correlation in {C}ryosat {SAR}
  echoes from ocean surfaces: implications for optimal pseudo-{LRM} waveform
  averaging,'' in \emph{Report of the Ocean Surf. Topogr. Sci. Team Meet.
  (OSTST)}, Venice-Lido, Italy, Sept. 22-29, 2012. [Online]. Available:
  \url{http://www.aviso.oceanobs.com/fileadmin/documents/OSTST/2012/oral/02_friday_28/05_instr_processing_IIb/03_IP2B_Smith1.pdf}
\BIBentrySTDinterwordspacing

\bibitem{Gommenginger2012OSTST}
\BIBentryALTinterwordspacing
C.~Gommenginger, P.~Cipollini, D.~Cotton, S.~Dinardo, and J.~Benveniste,
  ``Finer, better, closer: Advanced capabilities of {SAR} altimetry in the open
  ocean and the coastal zone,'' in \emph{Report of the Ocean Surf. Topogr. Sci.
  Team Meet. (OSTST)}, Venice-Lido, Italy, Sept. 22-29, 2012. [Online].
  Available:
  \url{http://www.aviso.oceanobs.com/fileadmin/documents/OSTST/2012/oral/02_friday_28/02_instr_processing_II/01_IP2_Gommenginger.pdf}
\BIBentrySTDinterwordspacing

\bibitem{Boy2012}
\BIBentryALTinterwordspacing
F.~Boy, J.-D. Desjonqueres, A.~Halimi, P.~Thibaut, T.~Moreau, and N.~Picot,
  ``{C}ryosat {P}rocessing {P}rototype, {LRM} and {SAR} processing on {CNES}
  side and a comparison to {DUACS SLA},'' in \emph{Report of the Ocean Surf.
  Topogr. Sci. Team Meet. (OSTST)}, Venice-Lido, Italy, Sept. 22-29, 2012.
  [Online]. Available:
  \url{http://www.aviso.oceanobs.com/fileadmin/documents/OSTST/2012/oral/02_friday_28/02_instr_processing_II/02_IP2_Boy.pdf}
\BIBentrySTDinterwordspacing

\bibitem{Puig2008}
C.~Martin-Puig, G.~Ruffini, J.~Marquez, D.~Cotton, M.~Srokosz, P.~Challenor,
  R.~K. Raney, and J.~Benveniste, ``Theoretical model of {SAR} altimeter over
  water surfaces,'' in \emph{Proc. IEEE Int. Conf. Geosci. Remote Sens.
  (IGARSS)}, Boston, U.S.A., July 7-11, 2008, pp. 242 -- 245.

\bibitem{Gommenginger2012}
C.~Gommenginger, C.~Martin-Puig, S.~Dinardo, D.~Cotton, and J.~Benveniste,
  ``Improved altimetric performance of {C}ryosat-2 {SAR} mode over the open
  ocean and the coastal zone,'' in \emph{Proc. IEEE Int. Conf. Geosci. Remote
  Sens. (IGARSS)}, Munich, Germany, July 22-27, 2012.

\bibitem{Gomez2009}
J.~G\'omez-Enri, S.~Vignudelli, G.~D. Quartly, C.~P. Gommenginger,
  P.~Cipollini, P.~G. Challenor, and J.~Benveniste, ``Modeling {ENVISAT RA}-2
  waveforms in the coastal zone: Case study of calm water contamination,''
  \emph{IEEE Geosci. Remote Sens. Lett.}, vol.~7, no.~3, pp. 474--478, July
  2010.

\bibitem{Robert2007}
C.~P. Robert, \emph{The Bayesian Choice: from Decision-Theoretic Motivations to
  Computational Implementation}, 2nd~ed., ser. Springer Texts in
  Statistics.\hskip 1em plus 0.5em minus 0.4em\relax New York: Springer-Verlag,
  2007.

\bibitem{Halimi_ArXiv2014}
A.~Halimi, N.~Dobigeon, and J.-Y. Tourneret, ``Unsupervised unmixing of
  hyperspectral images accounting for endmember variability,'' in \emph{ArXiv
  e-prints}, Jun. 2014.

\end{thebibliography}
\end{document}